\documentclass[twocolumn,10pt]{article}   	
\usepackage{geometry}                		
\usepackage{graphicx}				
\usepackage{amssymb}

\title{\bf A heuristic account of the radiation by the superluminally moving current sheet in the magnetosphere of a neutron star}
\author{Houshang Ardavan\footnote{ardavan@ast.cam.ac.uk}\\
Institute of Astronomy, University of Cambridge,\\
Madingley Road, Cambridge CB3 0HA, UK}
\date{}							

\begin{document}
\maketitle

\begin{abstract}
Results of the mathematical treatment of the radiation by the superluminally moving current sheet in the magnetosphere of a neutron star, which was presented in Ardavan (2021, {\it MNRAS}, {\bf 507}, 4530), are explained here in more transparent physical terms with the aid of illustrations.  Not only do these results provide an all-encompassing explanation for the salient features of the radiation received from pulsars (its brightness temperature, polarization, spectrum, profile with microstructure and with a phase lag between the radio and gamma-ray peaks, and the discrepancy between the energetic requirements of its radio and gamma-ray components), but they also shed light on the putative energetic requirements of magnetars and the sources of fast radio bursts and gamma-ray bursts.
\end{abstract}

\section{Introduction}
\label{sec:Introduction}

The mechanism by which the radiation received from obliquely rotating neutron stars is generated remains an open question half a century after the discovery of pulsars~\cite{Beskin2018,Melrose2021}.  In contrast, considerable progress has recently been made in determining the structure of the magnetosphere that surrounds these objects: numerical computations based on the force-free, magnetohydrodynamic and particle-in-cell formalisms have now firmly established that the magnetosphere of an oblique rotator entails a current sheet outside its light cylinder whose rotating distribution pattern moves with linear speeds exceeding the speed of light in vacuum~\cite{SpitkovskyA:Oblique,Contopoulos:2012, Tchekhovskoy2013, Philippov2018}.  However, the role played by the superluminal motion of this current sheet in generating the multi-wavelength focused pulses of radiation that we receive from neutron stars is unknown.  Here we insert the description of the current sheet provided by the numerical simulations~\cite{Tchekhovskoy:etal} in the classical expression for the retarded potential and thereby determine the radiation field generated by this source in the time domain.  

Both the radiation field thus calculated and the electric and magnetic fields that pervade the pulsar magnetosphere are solutions of Maxwell's equations for the same charge-current distribution.  These two solutions are completely different, nevertheless,  because they satisfy different boundary conditions: the far-field boundary conditions with which the structure of the pulsar magnetosphere is computed are radically different from the corresponding boundary conditions with which the retarded solution of these equations (i.e. the solution describing the radiation from the charges and currents in the pulsar magnetosphere) is derived (see Section~3 and the last paragraph in Section~6 of~\cite{Ardavan2021}).  

Given that the superluminally moving distribution pattern of the magnetospheric current sheet is created by the coordinated motion of aggregates of subluminally moving charged particles (see~\cite{GinzburgVL:vaveaa, BolotovskiiBM:VaveaD,BolotovskiiBM:Radbcm}), the motion of any of its constituent particles is too complicated to be taken into account individually.  Only the densities of charges and currents enter the Maxwell's equations, on the other hand, so that the macroscopic charge-current distribution associated with the magnetospheric current sheet takes full account of the contributions toward the radiation that arise from the complicated motions of the charged particles comprising it.   

\begin{figure}
\centerline{\includegraphics[width=12cm]{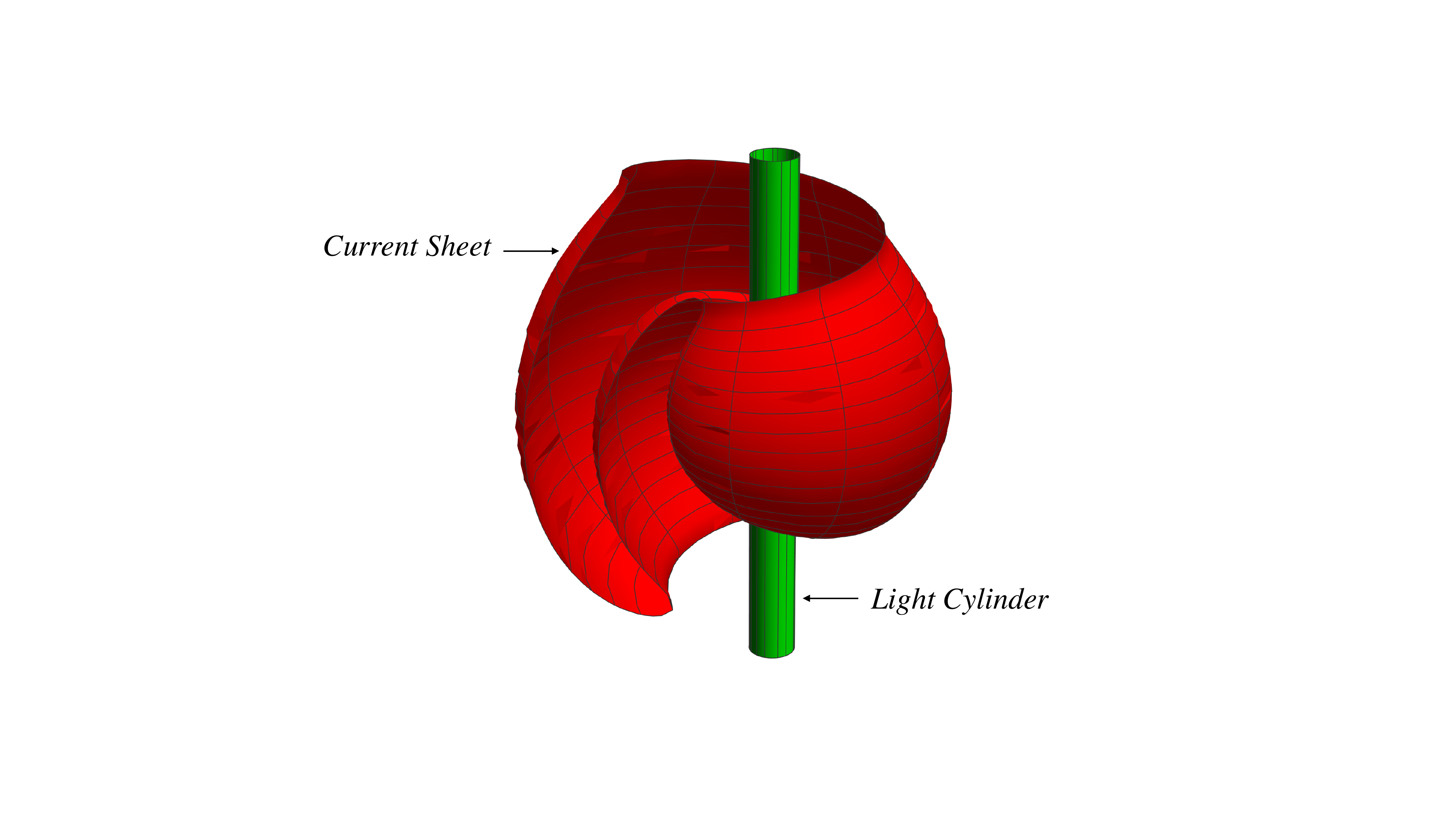}}
\caption{Snapshot of a single turn of the current sheet about the light cylinder (i.e.\ $r=c/\omega$, where $r$ is the distance from the rotation axis, $\omega$ is the angular frequency of rotation of the central neutron star and $c$ is the speed of light in vacuum) for the value $\alpha=\pi/3$ of the angle between the the magnetic and rotation axes of the star.  This surface undulates within the latitudinal interval $\pi/2-\alpha\le\theta\le\pi/2+\alpha$ each time it turns about the rotation axis, thus wrapping itself around the light cylinder as it extends to the outer edge of the magnetosphere.}
\label{FH0}
\end{figure}

We find that the radiation generated by the current sheet consists of highly focused pulses whose salient features (brightness temperature, polarization, spectrum, profile with microstructure and with a phase lag between the radio and gamma-ray peaks, and the discrepancy between the energetic requirements of its radio and gamma-ray components) are strikingly similar to those of the emission received from pulsars~\cite{Ardavan2021,Ardavan2024Radio,Ardavan2023Crab,ArdavanCVG,Ardavan2023}.  In certain latitudinal directions, the flux density of this radiation diminishes with the distance $D$ from the star as $D^{-3/2}$ (rather than $D^{-2}$) over specific frequency bands: bands whose central frequencies range from those of radio waves to gamma-rays depending on the thickness of the current sheet.  This result suggests that the high energetic requirements normally attributed to gamma-ray pulsars and magnetars and to the sources of fast radio bursts and gamma-ray bursts could be artefacts of the assumption that the radiation fields of all sources necessarily decay as predicted by the inverse-square law~\cite{Ardavan2023,Ardavan_Fermi, Ardavan_magnetar}.

The surface on which the magnetospheric current sheet is distributed (Fig.~\ref{FH0}) spirals away from the light cylinder in the azimuthal direction at the same time as undulating in the latitudinal direction~\cite{Tchekhovskoy:etal, Bogovalov1999}.  Its motion consists of a rigid rotation with the same angular frequency, $\omega$, as that of the central neutron star and a radial expansion (resulting from the combination of its spiral structure and rigid rotation) with the speed of light in vacuum, $c$.  This is not incompatible with the requirements of special relativity because the superluminally moving distribution pattern of the current sheet is created by the coordinated motion of aggregates of subluminally moving charged particles~\cite{GinzburgVL:vaveaa, BolotovskiiBM:VaveaD, BolotovskiiBM:Radbcm}.

In this paper we treat the distribution of charges and currents that make up the current sheet at any given time as a prescribed volume source whose density can be inserted in the retarded solution of the inhomogeneous Maxwell's equations to find the radiation field it generates in unbounded free space.  The only role we assign to the rest of the magnetosphere, whose radiation field is negligibly weaker than that of the current sheet, is to maintain the propagation of this sheet.  The multi-wavelength focused pulses emitted by the current sheet escape the plasma surrounding the neutron star in the same way that the radiation generated by the accelerating charged particles invoked in most current attempts at modelling the emission mechanism of these objects does~\cite{Philippov2018,Philippov2019}.  

\begin{figure*}
\centerline{\includegraphics[width=12cm]{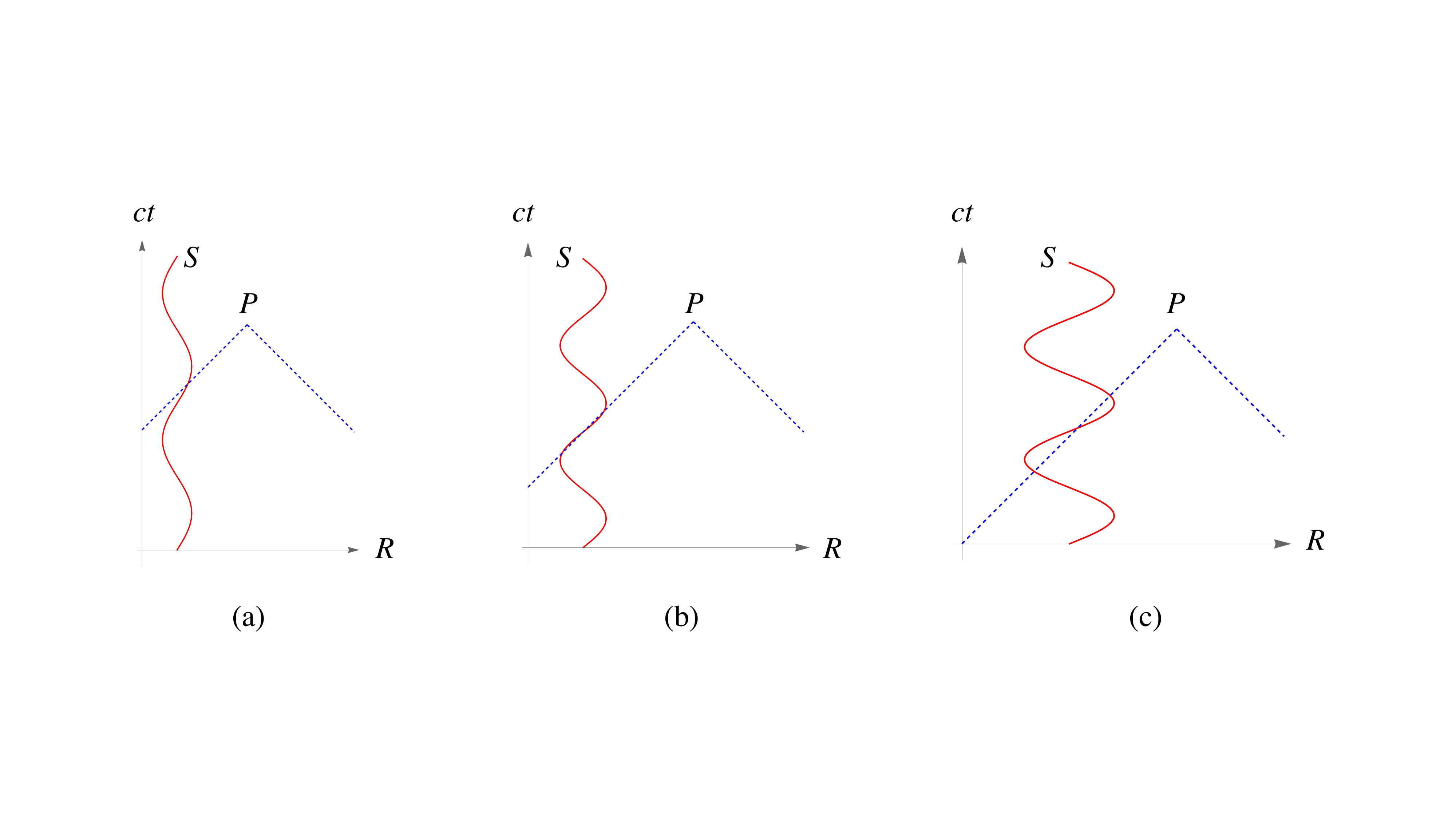}}
\caption{Space-time depictions of the past light cone of the observation point $P$ located at ${\bf x}_P$ (the blue dashed lines) and the possible forms of the trajectory, $R(t)=\vert{\bf x}(t)-{\bf x}_P\vert$, of a superluminally rotating source point $S$ with the position vector ${\bf x}$ (the red curves).  In (a), where $\vert{\rm d}R/{\rm d}t\vert<c$, the trajectory of $S$ intersects the past light cone of any $P$ only once, so that the field observed at $P$ at time $t_P$ is determined by the single wave front that was emitted by $S$ at the retarded time $t$.  In (c), where the component of the velocity of $S$ along the radiation direction is greater than $c$, the trajectory of $S$ intersects the past light cone of certain observers $P$ at three points, so that the field at $P$ receives contributions from three distinct values of the retarded time simultaneously. In (b), where both ${\rm d}R/{\rm d}t=-c$ and ${\rm d}^2R/{\rm d}t^2=0$ at the point where the trajectory of $S$ intersects the past light cone of $P$ tangentially, the field is determined by the coalescence of three contributions originating from a retarded time at which the source point approaches the observer along the radiation direction not only with the speed of light but also with zero acceleration.}
\label{FH1}
\end{figure*}

The current sheet is described by charge and current densities whose space-time distributions depend on the azimuthal coordinate $\varphi$ and time $t$ in the combination $\varphi-\omega t$ only~\cite{Tchekhovskoy:etal}.  The radiation field we are after can be built up, therefore, by the superposition of the fields of the uniformly rotating volume elements that constitute the distribution pattern of this source.  The radiation field of a rotating volume element whose linear speed exceeds the speed of light in vacuum receives simultaneous contributions from more than one retarded position of the source at certain observation points (Fig.~\ref{FH1}).  

\begin{figure*}
\centerline{\includegraphics[width=10cm]{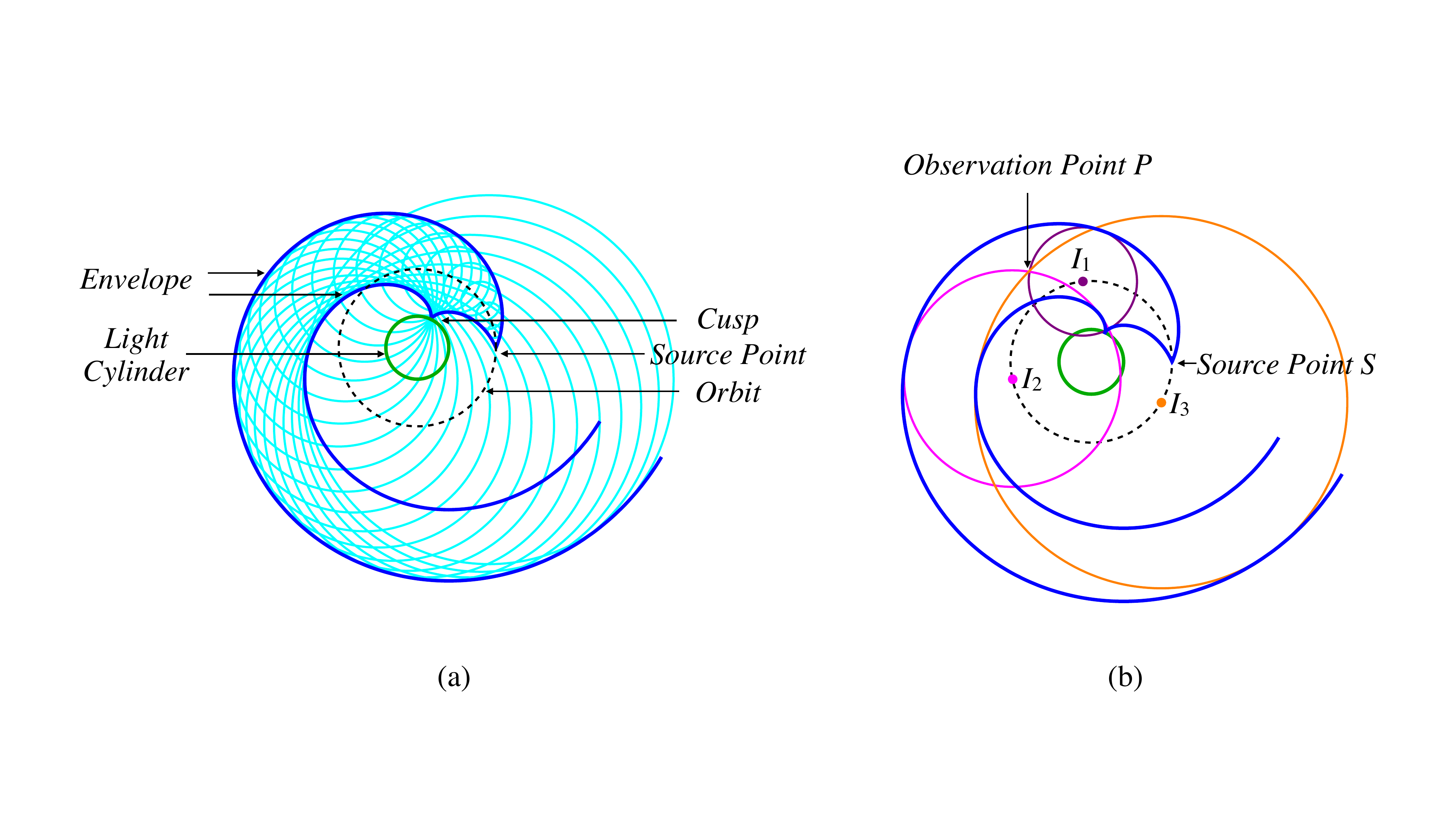}}
\caption{(a) Cross-sections of the wave fronts (the circles in light blue) emanating from a volume element of the current sheet with fixed radial and latitudinal coordinates.  This figure is plotted for a source element the radius of whose orbit (the dotted circle) is $2.5$ times the radius $c/\omega$ of the light cylinder (the green circle).  Cross-sections of the two sheets of the envelope of these wave fronts with the plane of the orbit (shown in dark blue) meet at a cusp and wind around the rotation axis, while moving away from it all the way to the far zone. (b) By receiving three wave fronts (the circles in brown, pink and orange) centred at three distinct retarded positions ($I_1$, $I_2$, $I_3$) of the source point $S$, an observer $P$ inside the envelope detects three images ($I_1$, $I_2$, $I_3$) of $S$ simultaneously.} 
\label{FH2}
\end{figure*}

\begin{figure*}
\centerline{\includegraphics[width=10cm]{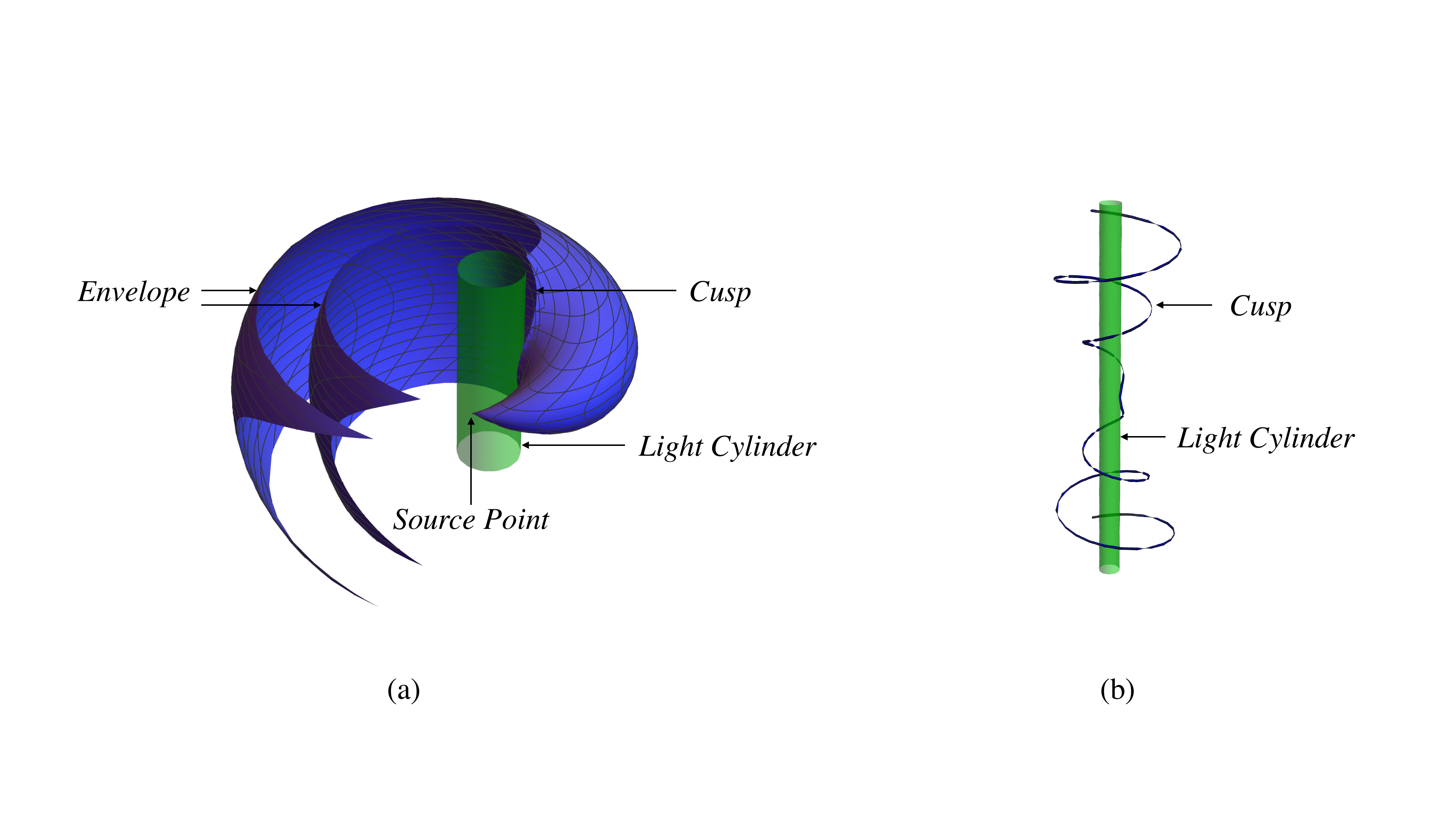}}
\caption{(a) Three-dimensional view of the two-sheeted envelope of wave fronts emanating from a uniformly rotating source element outside the light cylinder $r_P=c/\omega$, where $r_P$ is the distance of the observation point from the axis of roation.  This envelope spirals into the far zone maintaining a symmetry with respect to the plane of the orbit of the source element.  (b) The cusp locus along which the two sheets of the envelope meet.  This curve tangentially touches the light cylinder where it crosses the plane of the orbit and spirals into the far zone on a hyperboloid whose asymptotes lie along the cones $\theta_P=\arcsin[c/(r_{\rm S}\omega)]$ and $\theta_P=\pi-\arcsin[c/(r_{\rm S}\omega)]$, where $r_{\rm S}$ is the radius of the circular orbit of the source element.}
\label{FH3}
\end{figure*}

As a result, this field entails intersecting wave fronts that possess a two-sheeted cusped envelope (Figs.~\ref{FH2} and \ref{FH3}).  Outside the envelope only one wave front passes through the observation point at any given observation time; but inside the envelope three distinct wave fronts, emitted at three different values of the retarded time, simultaneously pass through each observation point (Section~\ref{sec:PointSource}).  Coalescence of two of the contributing retarded times on the envelope of wave fronts results in the divergence of the resulting field on this surface (reflecting the fact that a superluminally moving source cannot be point-like~\cite{GinzburgVL:vaveaa, BolotovskiiBM:VaveaD, BolotovskiiBM:Radbcm}).  At an observation point on the cusp locus of the envelope all three of the contributing retarded times coalesce and the field in question has a higher-order singularity.  This radiation field embraces a synergy between the vacuum version of the field of {\v C}erenkov radiation and the superluminal version of the field of synchrotron radiation. 

Constructive interference of the emitted waves and formation of caustics thus play a crucial role in determining the radiation field of the current sheet (Section~\ref{sec:volume}).  Its double-peaked pulse profiles and S-shaped polarization position angle distributions stem from the caustics associated with the nearby stationary points of certain phase functions.   What underpin the high brightness temperatures and the broad frequency spectra of this radiation are the extraordinary values of the amplitude and width of the pulses that are generated when the maxima and minima of the phase functions in question coalesce into inflection points.  There is always a latitudinal direction along which the separation between any pair of these nearby maxima and minima decreases with increasing distance from the source.  The enhanced focusing of the emitted waves that takes place as a result of the diminishing separation between the nearly coincident stationary points of a phase function gives rise, in turn, to a lower rate of decay of the flux density of the radiation with distance (Section~\ref{sec:charactersitics}).  That the decay of the present radiation with distance disobeys the inverse-square law in certain directions is not incompatible with the requirements of the conservation of energy because the radiation process discussed here is intrinsically transient: the difference in the fluxes of power across any two spheres centred on the star is balanced by the change with time of the energy contained inside the shell bounded by those spheres (see~\cite{Ardavan_JPP}, Appendix C).  

\section{The field generated by a constituent volume element of the source}
\label{sec:PointSource}

A superluminal source is necessarily volume-distributed~\cite{GinzburgVL:vaveaa, BolotovskiiBM:VaveaD, BolotovskiiBM:Radbcm}.  However, its field can be built up from the superposition of the fields of its constituent volume elements which are point-like.  The current sheet in the magnetosphere of a non-aligned neutron star that rotates with the angular frequency $\omega$ is described by charge and current densities outside the light cylinder whose distribution patterns rigidly rotate with the same angular frequency as that of the central neutron star~\cite{Tchekhovskoy:etal}.  The radiation field that is generated by this current sheet can be built up, therefore, by the superposition of the fields of uniformly rotating volume elements whose linear speeds $r\omega$ exceed $c$, where $r$ is the distance of a source point from the rotation axis.  

In the space-time $({\bf x}, t)$ of source points, the past light cone $\vert{\bf x}-{\bf x}_P\vert=c(t_P-t)$ of an observer $P$ with the space-time coordinates $({\bf x}_P, t_P)$ is a sphere centred on the observer's position ${\bf x}_P$ which collapses, with the speed of light $c$, onto the point ${\bf x}_P$ at the observation time $t_P$.  Since the waves it emits propagate with the speed $c$, a moving source point ${\bf x}={\bf x}(t)$ makes its contribution toward the radiation field detected at $({\bf x}_P, t_P)$ at the retarded (emission) time $t$ when this collapsing sphere crosses it, i.e.\ at the instant at which $R(t)\equiv\vert{\bf x}(t)-{\bf x}_P\vert=c(t_P-t)$.  The red curves in Fig.~\ref{FH1} depict the possible trajectories of a rotating source point in the $(R, ct)$ space.  The dashed lines in this figure represent the past light cone of the observer $P$.  Depending on their positions relative to the observer, various source points (volume elements) of a superluminally rotating source distribution approach the observer along the radiation direction $({\bf x}_P-{\bf x})/\vert{\bf x}_P-{\bf x}\vert$ with a speed ${\rm d}R/{\rm d}t$ at the retarded time (i.e.\ the time at which the space-time trajectory of the source element in question intersects the past light cone of the observer) that can be less than, equal to or greater than $c$ as in Figs.~\ref{FH1}a--\ref{FH1}c.  

The number of times the collapsing sphere $\vert{\bf x}-{\bf x}_P\vert=c(t_P-t)$ intersects a rotating source point at ${\bf x}(t)$ during the time it propagates past the circular orbit of such a point source depends on whether $\vert{\rm d}R/{\rm d}t\vert$ is less than, equal to or greater than $c$.  The trajectory of a source element for which $\vert{\rm d}R/{\rm d}t\vert<c$ intersects the past light cone of any observer only once and so makes its contribution toward the value of the observed radiation field at a single instant of retarded time (Fig.~\ref{FH1}a).  In contrast, a source element for which $\vert{\rm d}R/{\rm d}t\vert>c$ makes its contribution toward the value of the radiation field detected by certain observers at three distinct instants of the retarded time: a contribution comprising three distinct parts that are received simultaneously at the observation time $t_P$ (Fig.~\ref{FH1}c).  (For a sufficiently large $\vert{\rm d}R/{\rm d}t\vert$, the trajectory of the rotating source point can intersect the past light cone of the observer 5, 7, 9, \ldots times.)  The three parts of this contribution coalesce in the case of a source element that approaches the observer with the speed of light (${\rm d}R/{\rm d}t=-c$) and zero acceleration (${\rm d}^2R/{\rm d}t^2=0$) at the retarded time (Fig.~\ref{FH1}b). 

Figure~\ref{FH2}a shows the intersections with the equatorial plane of the spherical wave fronts emitted by a source element rotating with the constant angular frequency $\omega$ whose orbit lies outside the light cylinder $r_P=c/\omega$, i.e.\ whose speed exceeds $c$, where $r_P$ is the distance of the observation point from the rotation axis.  The envelope of these wave fronts consists of two sheets that meet at a cusp.  Only a single wave front propagates past an observer outside the envelope (for whom $\vert{\rm d}R/{\rm d}t\vert<c$) at any given time.  But an observer inside the envelope (for whom $\vert{\rm d}R/{\rm d}t\vert>c$) receives three intersecting wave fronts, emanating from three distinct retarded positions of the source element ($I_1$, $I_2$ and $I_3$), simultaneously (see Fig.~\ref{FH2}b).  On each sheet of the envelope, two of these wave fronts (together with the corresponding retarded times at which they are emitted) coalesce and interfere constructively to generate an infinitely large radiation field (\cite{Ardavan2021}, Section 3.4).  On the cusp locus of the envelope, where all three of these wave fronts (together with the corresponding images $I_1$, $I_2$ and $I_3$ of the source element) coalesce, the resulting radiation field has a higher-order singularity.  (Divergence of the field on the envelope and its cusp, which stem from the relativistic restrictions inherent in electrodynamics, reflect the fact that no superluminal source can be point-like~\cite{GinzburgVL:vaveaa, BolotovskiiBM:VaveaD, BolotovskiiBM:Radbcm}.)

Figure~\ref{FH3}a shows a three-dimensional view of the envelope of wave fronts that are emitted by a uniformly rotating source element outside the light cylinder.  The cusp along which the two sheets of this envelope meet touches the light cylinder on the plane of the orbit tangentially and spirals away from the rotation axis on a hyperboloid that is symmetrical with respect to that plane (see Fig.~\ref{FH3}b).  As in Fig.~\ref{FH2}, the number of wave fronts that pass the observation point at a given observation time are one, two or three outside, on or inside the envelope.  Likewise, the generated radiation is most tightly focused on the cusp curve shown in Fig.~\ref{FH3}b where the waves from three coalescent retarded positions (images) of the source element interfere constructively.

\begin{figure*}
\centerline{\includegraphics[width=10cm]{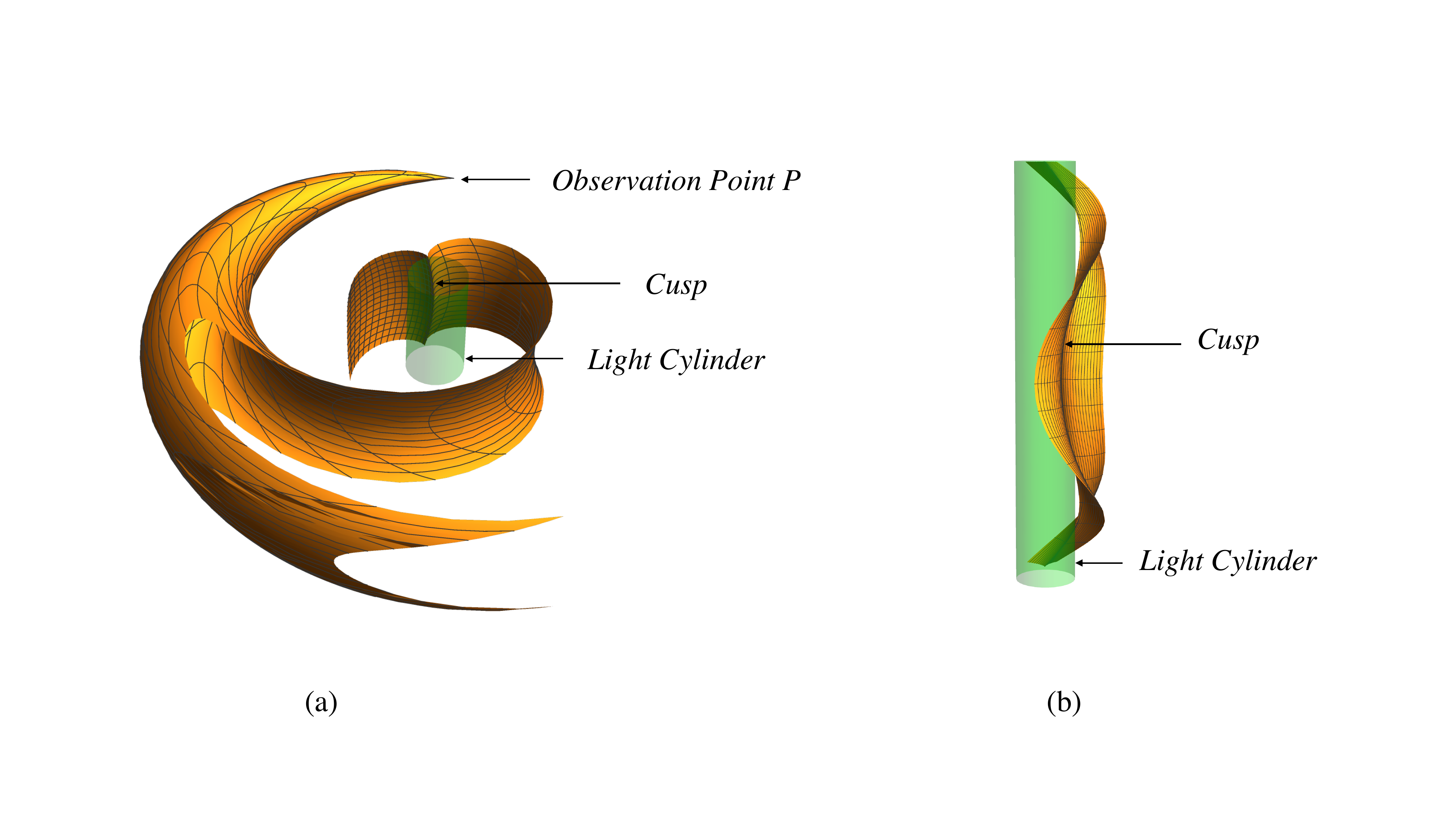}}
\caption{(a) The bifurcation surface of the observation point $P$, its cusp locus and the light cylinder $r=c/\omega$.  In contrast to the envelope of wave fronts shown in Fig.~\ref{FH3}a, which resides in the space of observation points, the surface shown here resides in the space of source points: it is the locus of source elements that approach $P$ along the radiation direction with the speed of light at the retarded time.  The two sheets of this surface meet along a cusp that tangentially touches the light cylinder where it crosses the plane of rotation containing the observation point and spirals outward as the colatitude $\theta$ decreases.  The source points on this cusp approach the observer along the radiation direction not only with the speed of light but also with zero acceleration at the retarded time.  (b) A close-up view of the two sheets of the bifurcation surface in the vicinity of the cusp locus of this surface.  The hyperboloid containing this cusp has the conical asymptotes $\theta=\arcsin[c/(r_P\omega)]$ and $\theta=\pi-\arcsin[c/(r_P\omega)]$.}
\label{FH4}
\end{figure*}

\section{The field generated by the entire volume of the source}
\label{sec:volume}

Depending on the speed $\vert{\rm d}R/{\rm d}t\vert$ with which they approach an observation point $P$ at the retarded time, different volume elements of an extended source make differing contributions toward the value of the radiation field at $P$ (see Fig.~\ref{FH1}).  To distinguish between the contributions from different sections of a volume source, we need to introduce the notion of bifurcation surface: a two-sheeted cusped surface reciprocal to the envelope of wave fronts that resides in the space of source points, instead of residing in the space of observation points, and issues from the observation point, instead of issuing from a source point (see Fig.~\ref{FH4}).  

Intersection of the bifurcation surface of an observation point with the volume of the source divides this volume into two parts.  The source elements inside the bifurcation surface make their contributions toward the observed field at three distinct values of the retarded time, while the source elements outside the bifurcation surface make their contributions at a single value of the retarded time (as a subluminally moving source would).  The source elements inside and close to the bifurcation surface (for which the values of two of the contributing retarded times approach one another) and the source elements inside the bifurcation surface close to its cusp (for which all three values of the contributing retarded times coalesce) are by far the dominant contributors toward the strength of the observed field.  

As noted before, the densities of charges and currents in the magnetosphere of a non-aligned neutron star that rotates with the constant angular frequency $\omega$ are functions of the azimuthal angle $\varphi$ and time $t$ in the combination ${\hat\varphi}=\varphi-\omega t$ only~\cite{Tchekhovskoy:etal}.  The coordinate ${\hat\varphi}$ labels each volume element of the uniformly rotating magnetosphere with its azimuthal position at the time $t=0$ (thus ranging over an interval of length $2\pi$).  To superpose the contributions from various volume elements of such a magnetosphere toward the value of the radiation field, we must perform the integration in the classical expression for the retarded field (\cite{Ardavan2021}, Section 3.2) over the volume occupied by the charges and currents in the $(r_{\rm s},{\hat\varphi}, \theta)$ space, where the radius $r_{\rm s}$ and the colatitude $\theta$ denote the corresponding spherical coordinates of these source elements.

\begin{figure*}
\centerline{\includegraphics[width=11cm]{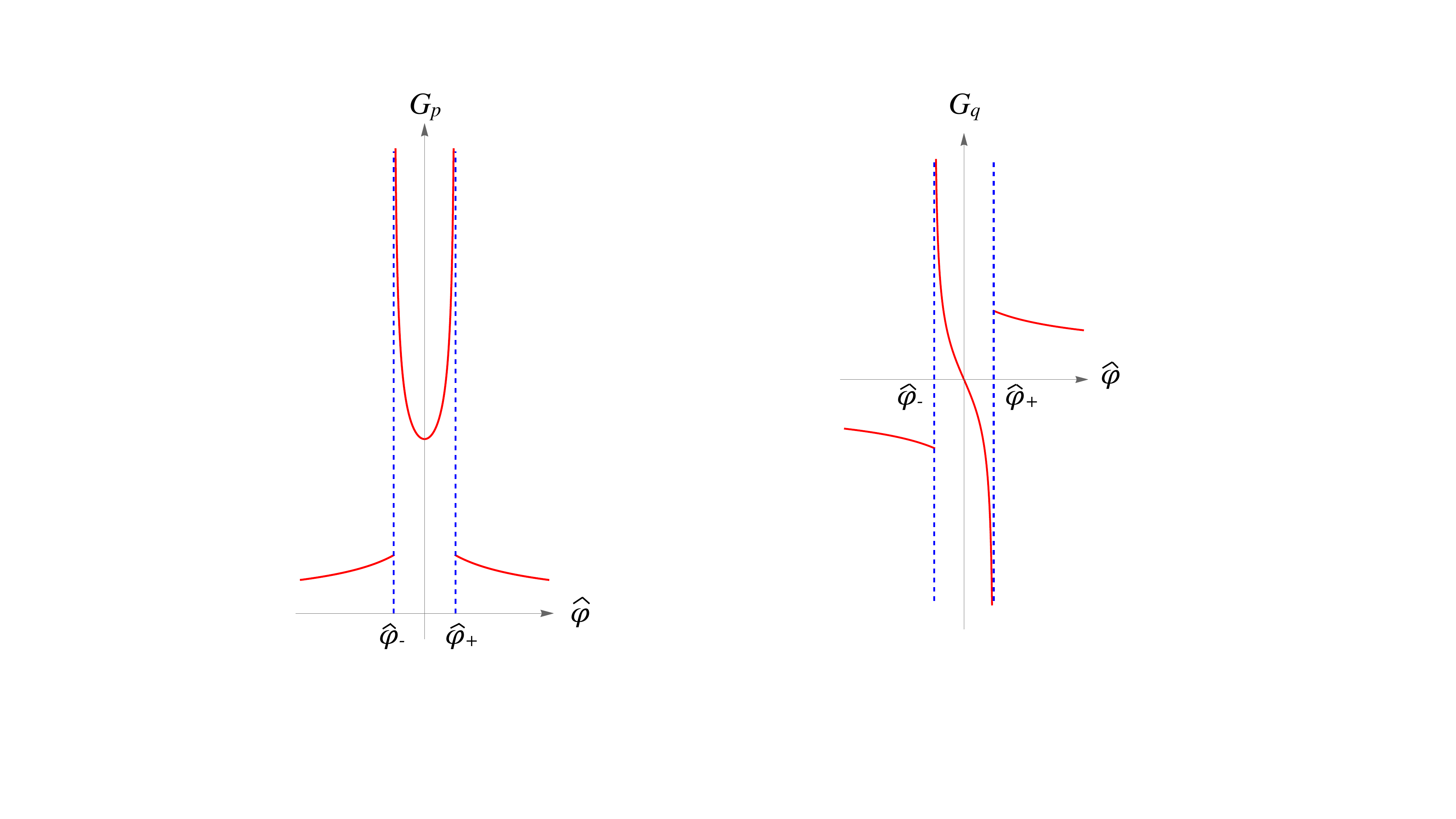}}
\caption{Schematic depiction of the ${\hat\varphi}$-dependence of the functions that describe the contribution, toward the observed field, of a superluminally rotating source element inside (${\hat\varphi}_-<{\hat\varphi}<{\hat\varphi}_+$) and outside (${\varphi}\le{\hat\varphi}_-$, ${\varphi}\ge{\hat\varphi}_+$) a cusped strip of the bifurcation surface (Fig.~\ref{FH4}b).  The contribution in question is described by a linear combination of these two functions.}
\label{FH5}
\end{figure*}

By virtue of approaching the observer at the retarded time with a speed that is close to $c$ and an acceleration that is vanishingly small, the source elements in the vicinity of the strip of the bifurcation surface that borders on its cusp (Fig.~\ref{FH4}b) make a contribution toward the observed radiation field that is singular on the inner and regular on the outer side of the bifurcation surface: the ${\hat\varphi}$-dependence of the contributions made by the elements inside (${\hat\varphi}_-<{\hat\varphi}<{\hat\varphi}_+$) and outside (${\hat\varphi}\le{\hat\varphi}_-$, ${\hat\varphi}\ge{\hat\varphi}_+$) a cusped strip of the bifurcation surface (Fig.~\ref{FH4}b) is given by a linear combination of the two functions shown in Fig.~\ref{FH5}.  The contribution of each constituent ring ($r_{\rm s}=$ const., $\theta=$ const., $0\le{\hat\varphi}<2\pi$) of a superluminally rotating source distribution toward the field detected at an observation point $P$ is thus discontinuous across the two sheets of the bifurcation surface associated with $P$ (\cite{Ardavan2021}, Section 3.3).

Superposition of the contributions of the source elements that constitute a rotating ring ($r_{\rm s}=$ const., $\theta=$ const., $0\le{\hat\varphi}<2\pi$) toward the field (i.e.\ an integration of the classical expression for the retarded field with respect to ${\hat\varphi}$ at fixed values of $r_{\rm s}$ and $\theta$) shows that, in addition to a contribution identical to the conventional one which is obtained in the case of a subluminal source, the present radiation field receives contributions from the discontinuities (depicted in Fig.~\ref{FH5}) across the bifurcation surface of the observation point (\cite{Ardavan2021}, Section 3.5).  These additional contributions, as we shall see below, turn out to result in an unconventional radiation field, much stronger than that resulting from the conventional contribution, whose characteristics differ from those of any previously known radiation field (\cite{Ardavan2021}, Section 4.1).

Next, we superpose the fields of the constituent rotating rings of the source distribution that lie on a cone outside the light cylinder, i.e.\ the rings for which $\theta$ is constant and $r_s$ ranges from $c/(\omega\sin\theta)$ to the outer edge of the magnetosphere.  The dominant contribution toward the value of the field of this truncated cone at the observation point $P$ happens to come from those source elements of it that lie in the vicinity of the cusp locus of the bifurcation surface of $P$, i.e.\ the elements that approach the observation point along the radiation direction with speeds that are close to $c$ and accelerations that are vanishingly small at the retarded time (\cite{Ardavan2021}, Section 4.2).  This is not an unexpected result, given that the waves that emanate from such elements are focused most tightly (see Section~\ref{sec:PointSource}).

\begin{figure*}
\centerline{\includegraphics[width=11cm]{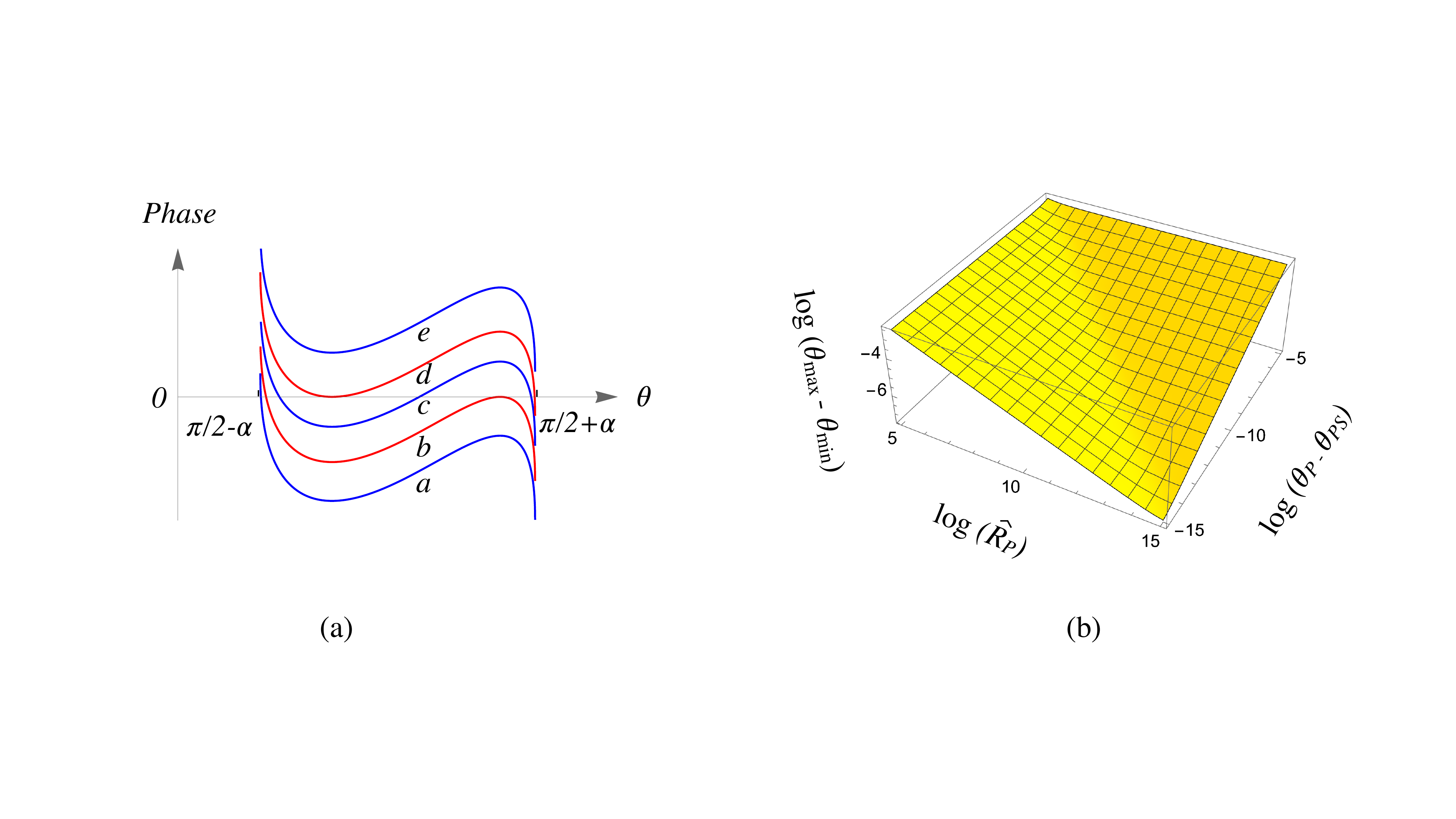}}
\caption{(a) Dependence on colatitude $\theta$ of the phase of one of the oscillating exponentials that are encountered after the contributions, toward the radiation field, of the constituent rotating rings of the source distribution having the same colatitude but differing radii are superposed (see Section~\ref{sec:volume}).  The curves $a$--$e$ correspond to successively increasing values of the azimuthal coordinate $\varphi_P$ of the observation point.  Note that in the cases of the curves $b$ and $d$ (shown in red) this phase vanishes at one of its turning points.  (b) Dependence of the separation $\theta_{\rm max}-\theta_{\rm min}$ between the locations of the maximum and minimum of a phase with nearby turning points on the distance ${\hat R}_P$ of the observer from the source and on the departure $\theta_P-\theta_{PS}$ of the colatitude of the observer $\theta_P$ from the critical colatitude $\theta_{PS}$ at which the maximum and minimum of such a phase coalesce.  This figure shows that $\theta_{\rm max}-\theta_{\rm min}$ decreases as ${\hat R}_P^{-1/2}$ with increasing distance when $(\theta_P-\theta_{PS})\lesssim{\hat R}_P^{-1}$.}
\label{FH6}
\end{figure*}

The resulting field of the aggregate of superluminally rotating source elements that constitute the truncated cone in question is described by a linear combination of four oscillating exponentials whose phases are proportional to the radiation frequency.  As functions of the colatitude $\theta$ of the source elements, the phases of these exponential factors either have two turning points (as shown in Fig.~\ref{FH6}a) or vary monotonically (\cite{Ardavan2021}, Section 4.4).  For radiation frequencies that are appreciably larger than the rotation frequency $\omega/2\pi$, therefore, the dominant contribution toward the value of the field of the entire source (i.e.\ the field that results from the superposition of the fields of the truncated cones with differing opening angles $\theta$) is expected to arise from the neighbourhoods of those values of $\theta$ at which the phases of one or more of these oscillating exponentials have turning points: the rapid oscillations (between positive and negative values) of these exponential factors as functions of $\theta$ bring about the cancellation of the contributions from all other values of $\theta$ when the fields of the truncated cones that make up the entire source are superposed (\cite{Ardavan2021}, Section 4.5).

Changes in the value of the azimuthal coordinate $\varphi_P$ of the observation point simply shift the curve representing the $\theta$-dependence of the phase of a given exponential factor up or down without altering its shape: the curves $a$--$e$ in Fig.~\ref{FH6}a correspond to successively increasing values of the longitude $\varphi_P$ across the window occupied by the emitted signal.  In the case of the phase function depicted in Fig.~\ref{FH6}a, the amplitude of the emitted signal peaks at the two longitudes at which a stationary point of this phase function crosses the horizontal axis, i.e.\ at the values of $\varphi_P$ for which this phase function vanishes at its maximum or minimum (as do the curves labelled $b$ and $d$ in Fig.~\ref{FH6}a).  Depending on the fraction of the four oscillating exponentials whose phases have turning points, the number of peaks of the emitted signal can thus range from 2 to 8 (\cite{Ardavan2021}, Section 4.4).

For any given inclination angle $\alpha$, there are a number (between $2$ to $8$) of critical values of the colatitude $\theta_P$ of the observation point (denoted as $\theta_{PS}$) at which the maximum and minimum of the phase of one or more of the exponentials that have two turning points coincide and form an inflection point.  Close to the critical colatitude $\theta_{PS}$, the separation $\theta_{\rm max}-\theta_{\rm min}$ between the locations of the maximum and minimum of such phases has the dependence shown in Fig.~\ref{FH6}b on the departure of the colatitude of the observer from its critical value, $\theta_P-\theta_{PS}$, and on the distance ${\hat R}_P$ of the observer from the source in units of the light-cylinder radius.  It can be seen from Fig.~\ref{FH6}b that the separation between the maximum and minimum of the phase shown in Fig.~\ref{FH6}a decreases as ${\hat R}_P^{-1/2}$ with increasing distance of the observer from the source when $(\theta_P-\theta_{PS})\lesssim{\hat R}_P^{-1}$: a result that implies that the emitted signal gets more focused the further away it is from its source (\cite{Ardavan2021}, Section 5.5). 

If we now attempt to integrate the resulting expression for the field of the entire source over all frequencies, to obtain the time-domain description of this field, we encounter a divergent integral.  This divergence stems from the fact that, by entailing a step-like discontinuity in the magnetic field~\cite{Tchekhovskoy:etal}, the numerical simulations of the magnetosphere assign a zero width to the current sheet.  The vanishing thickness of the current sheet in turn results in an infinitely broad spectral distribution for the generated radiation field.  But, given that it is created by the coordinated motion of aggregates of subluminally moving particles, a superluminally moving source is necessarily volume-distributed: it can neither be point-like nor be distributed over a line or a surface~\cite{GinzburgVL:vaveaa, BolotovskiiBM:VaveaD, BolotovskiiBM:Radbcm}.  

In a physically more realistic model of the magnetosphere, where the processes that occur on plasma scales are taken into account, the curent sheet would have a non-zero thickness and the divergence in question would not occur.  We have removed the singularity that arises from overlooking the finite width of the current sheet by setting a lower limit, $c/(\kappa_{\rm u}\omega)$, on the wavelength of fluctuations associated with the microstructure of the pulse profile and treating $\kappa_{\rm u}$ as a free parameter.  Thickness of the current sheet is dictated by microphysical processes that are not well understood: the standard Harris solution of the Vlasov-Maxwell equations which is commonly used in analysing a current sheet is not applicable in the present case because the current sheet in the magnetosphere of a neutron star moves faster than light and so has no rest frame.  Even in stationary or subluminally moving cases, there is no consensus on whether equilibrium current sheets in realistic geometries have finite or zero thickness~\cite{Klimchuk}.  The upper limit on frequency of the radiation is, in our treatment, proportional to the free parameter $\kappa_{\rm u}$ with a proportionality factor that is larger the closer are the turning points of the phase function(s) with nearly coincident stationary points which is (are) responsible for generating the high-frequency radiation.  It turns out that the time-domain description of the radiation field, for large $\kappa_{\rm u}$, can be derived explicitly (\cite{Ardavan2021}, Section 4.7).

\begin{figure*}
\centerline{\includegraphics[width=11cm]{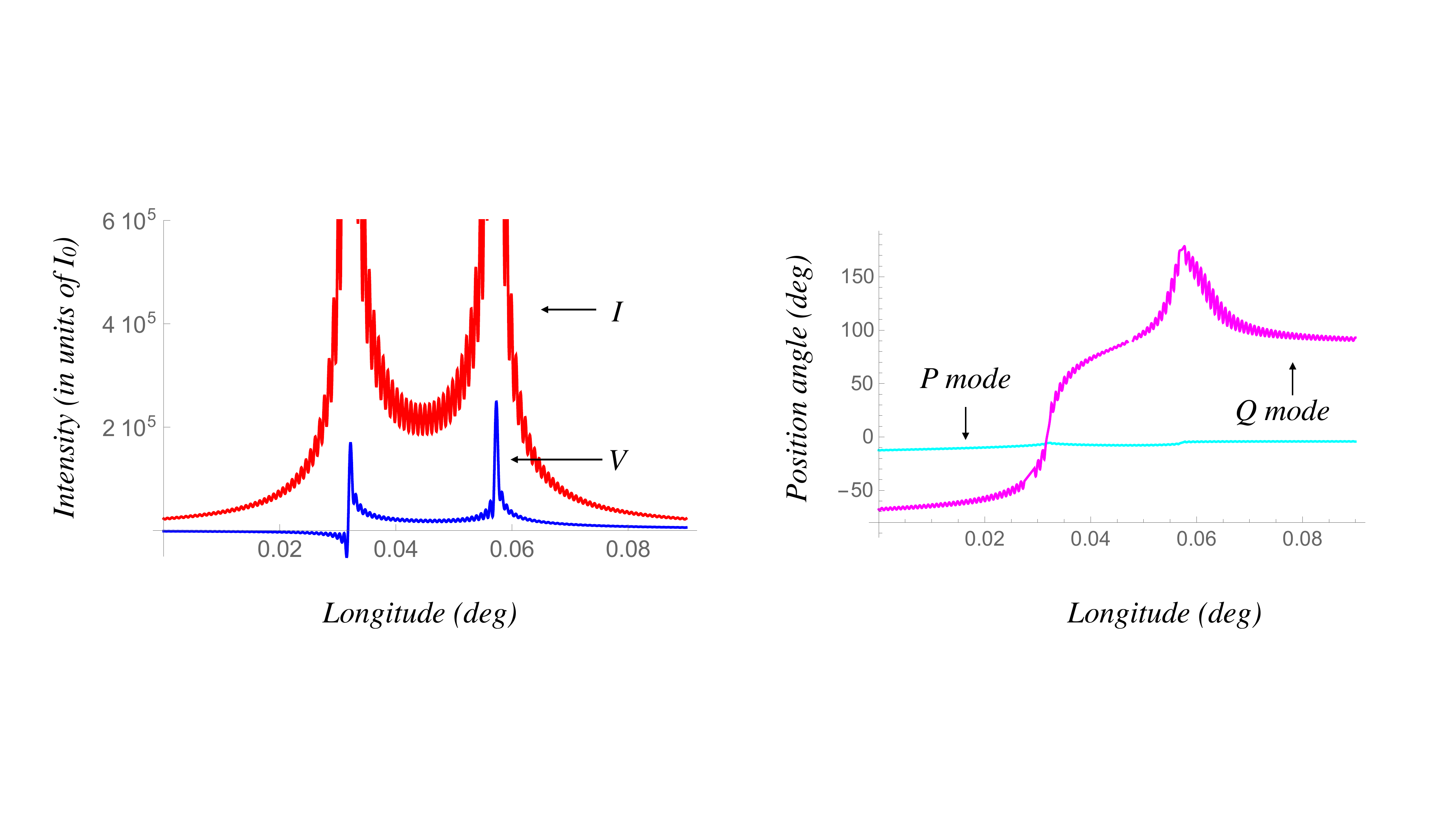}}
\caption{The Stokes parameters $I$, $V$ and the position angles $\psi$ of the polarization modes $P$ and $Q$ at an observation point with the colatitude $\theta_P=5^\circ$ for the inclination angle $\alpha=5^\circ$ and $\kappa_{\rm u}=10^2$.  Distribution of the Stokes parameter $L$ is essentially coincident with that of $I$.  The large value of the intensity and the short duration of this pulse stem from the proximity of the maximum and minimum of one of the phase functions described in Section~\ref{sec:volume}.  At its peak, the right-hand component of this pulse has the intensity $I=1.05\times10^{12}I_0$ and the longitudinal width $6.76\times{10^{-9}}$ deg when $\kappa_{\rm u}$ is $10^7$.  Note the reversal of the sense of circular polarization and the swing through $180^\circ$ of the position angle of the $Q$ polarization mode across the left-hand component of the pulse profile and the orthogonality of the position angles of the two polarization modes over the right-hand component of the pulse profile.  This is the only pulse occurring in the entire pulse window.  Here and in Figs.~\ref{FH8}--\ref{FH16} the origin of the azimuthal coordinate $\varphi_P$ of the observation point is shifted to place the starting point of the plotted pulse profile at longitude zero.  Unless stated otherwise, the value of ${\hat R}_P$ used for plotting the various pulse profiles is $10^6$.}
\label{FH7}
\end{figure*}

\section{Characteristics of the resulting radiation}
\label{sec:charactersitics}

The results described in this Section are consequences mainly of the shape and motion of the current sheet: two features of the magnetosphere which are the same not only for a wide class of magnetic-field configurations and plasma types~\cite{Bogovalov1999, Beskin2018}, but also both close to and far from the light cylinder.  As outlined in Sections~\ref{sec:PointSource} and~\ref{sec:volume}, these results are found by inserting the charge and current densities that follow from the semi-analytic description of the magnetosphere~\cite{Tchekhovskoy:etal} in the classical expression for the retarded potential, thereby evaluating the radiation field in the time domain~\cite{Ardavan2021}.  The characteristics of the resulting radiation described below, i.e., its pulse profile and polarization position-angle distribution, its brightness temperature, its frequency spectrum, the phase lag between its low-frequency and high-frequency peaks, and the distance dependence of its flux density, are then obtained with the aid of its Stokes parameters and Fourier representation.

\subsection{Pulse profiles and polarization position angles}
\label{subsec:profiles}

We begin by plotting the Stokes parameters $I$, $V$ and $L$ (i.e.\ the intensities, respectively, of the total, circularly and linearly polarized parts) of the resulting radiation in units of 
\begin{equation}
I_0=\left(\frac{B_0{\hat r}_{s0}^2}{{\hat R}_P}\right)^2
\label{E1}
\end{equation}
as functions of the longitude $\varphi_P$ of a far-field observation point $P$ for various values of its colatitude $\theta_P$, the inclination angle $\alpha$ and the lower bound $c/(\kappa_{\rm u}\omega)$ on the wavelength of the modulations of the pulse profile, where $B_0$ is the magnitude of the star's dipolar field at its magnetic pole and ${\hat r}_{s0}$ and ${\hat R}_P$ are the radius of the star and the distance of the observer from the source in units of the light-cylinder radius $c/\omega$.

The number of components of a pulse profile is determined by the number of stationary points of the four phase functions described in Section~\ref{sec:volume}.  The longitudinal positions of various components of a pulse profile are determined by the azimuthal coordinate $\varphi_P$ of the observation point for which these phase functions vanish at their maxima or minima.  To be able to depict the pulse profiles over the entire longitudinal intervals occupied by their various components while displaying the finite width of each component, we will plot the profiles of high-intensity pulses for $\kappa_{\rm u}=10^2$.  It should be borne in mind, however, that the value of intensity at the peak of each of the shown pulse components is a linearly increasing function of $\kappa_{\rm u}$.  Moreover, the higher the value of $\kappa_{\rm u}$, the narrower and closer to one another are the rapid low-amplitude modulations (i.e.\ the microstructure) of the pulse profile.  For values of $\kappa_{\rm u}$ higher than $10^4$, these modulations are too sharp and dense to show up in most of the figures plotted here.  Furthermore, we will adopt those branches of  the multi-valued function arctan appearing in the definition of the polarization position angle $\psi$ (in Eq.\ 170 of~\cite{Ardavan2021}) that yield a continuous distribution of this angle across various components of a given pulse. 

\begin{figure*}
\centerline{\includegraphics[width=12cm]{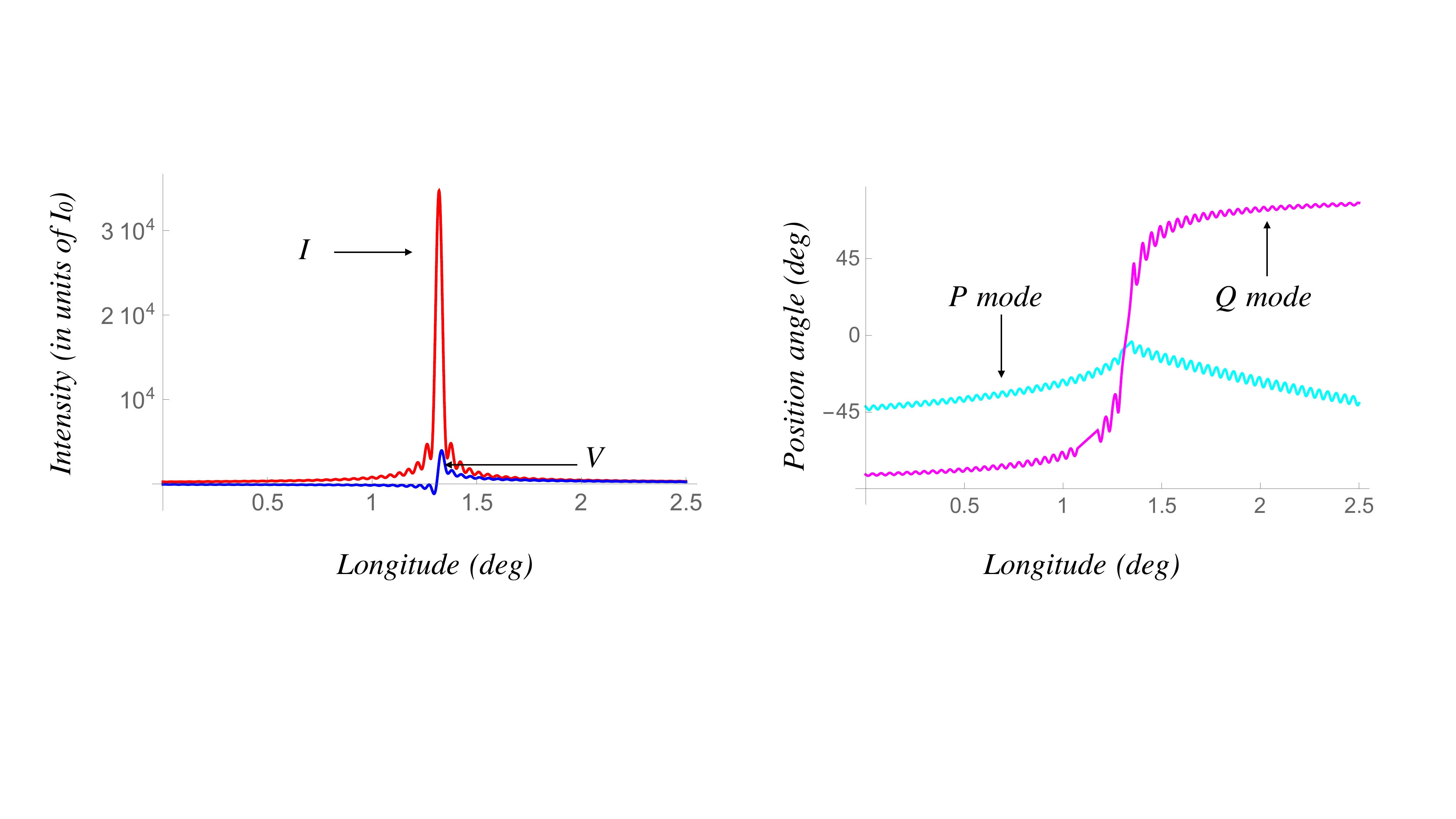}}
\caption{The Stokes parameters $I$, $V$ and the position angles $\psi$ of the polarization modes $P$ and $Q$ at an observation point with the colatitude $\theta_P=2.5^\circ$ for the inclination angle $\alpha=5^\circ$ and $\kappa_{\rm u}=10^4$.  Note the S-shaped swing through $180^\circ$ of the position angle of the $Q$ polarization mode across the pulse profile and the reversal of the sense of circular polarization across the peak of the pulse.  The pulse window encompasses another similar pulse at a longitudinal distance of about $140^\circ$ from this one.}
\label{FH8}
\end{figure*} 

\begin{figure*}
\centerline{\includegraphics[width=11cm]{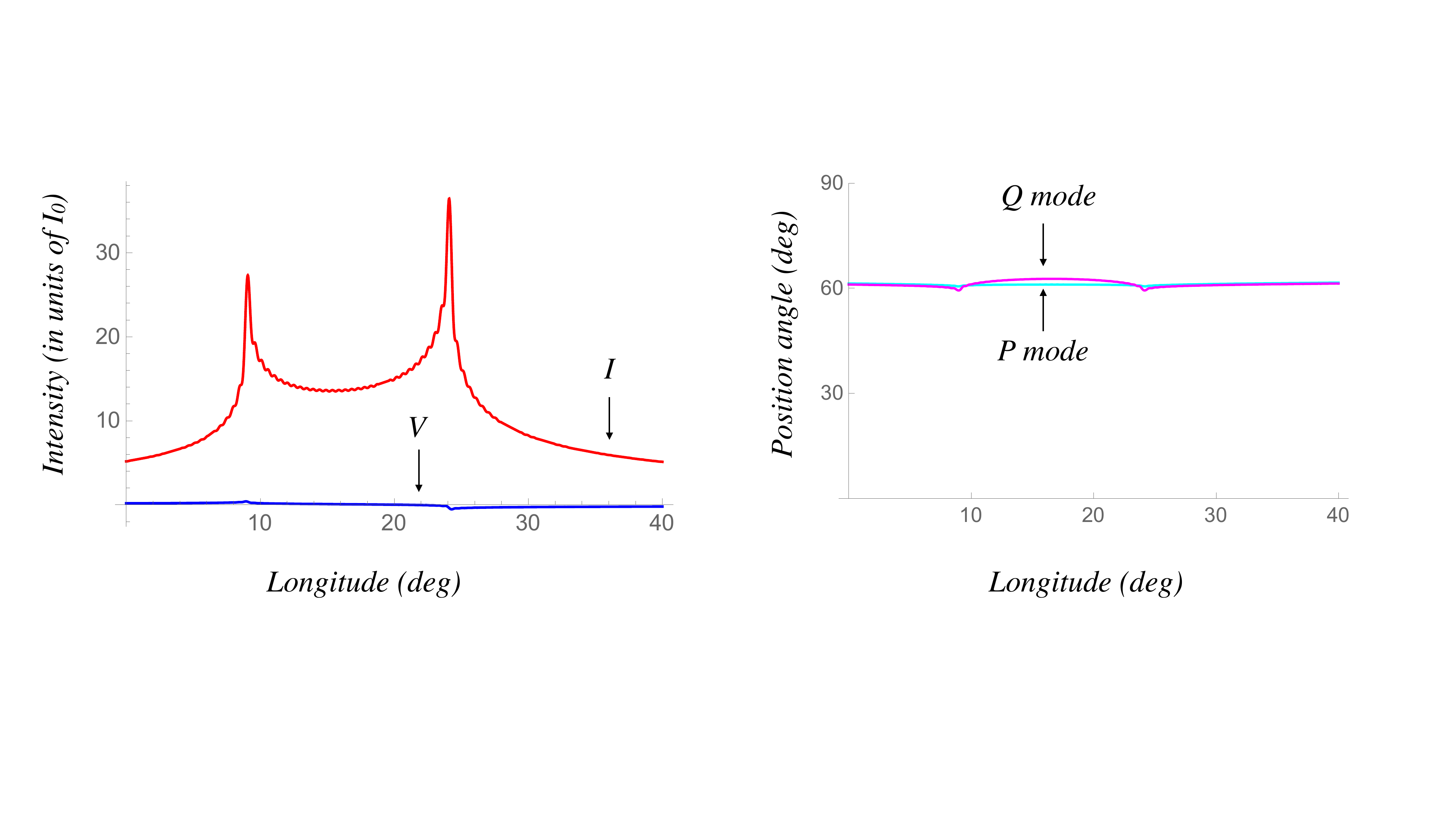}}
\caption{The Stokes parameters $I$, $V$ and the position angles $\psi$ of the polarization modes $P$ and $Q$ at an observation point with the colatitude $\theta_P=20^\circ$ for the inclination angle $\alpha=5^\circ$ and $\kappa_{\rm u}=10^2$.  In this example, the phase of only one of the oscillating exponentials described in Section~\ref{sec:volume} has stationary points.  The two peaks of the pulse occur where the phase in question vanishes at its maximum and minimum (see Fig.~\ref{FH6}a).  Though hardly discernible, the circular polarization of this pulse reverses sense across its both peaks.  This is the only pulse occurring in the entire pulse window.}
\label{FH9}
\end{figure*}

The examples in Figs.~\ref{FH7}--\ref{FH15} illustrate the longitudinal distributions of the obtained pulse profiles for a range of values of the inclination angle $\alpha$ and observer's colatitude $\theta_P$.  The electric field of the radiation generated by the current sheet turns out to be the sum of two distinct parts with differing polarization position angles.   Figures~\ref{FH7}--\ref{FH15} also show the longitudinal distributions of the position angles of these two parts which we refer to as $P$ and $Q$ modes.  Depending on the values of $\alpha$ and $\theta_P$, which determine the number and positions of the stationary points of various phase functions, the pulse profiles can be peaked singly ({Fig.~\ref{FH8}), doubly (Figs.~\ref{FH7}, \ref{FH9}--\ref{FH12}) or multiply (Figs.~\ref{FH13}--\ref{FH15}), can have both narrow (Figs.~\ref{FH7}, \ref{FH8}, \ref{FH12} and \ref{FH15}) and wide (Figs.~\ref{FH11}, \ref{FH13} and \ref{FH14}) widths and can comprise components that lie in either one (Figs.~\ref{FH7}, \ref{FH9} and \ref{FH11}) or two widely separated (Figs.~\ref{FH8}, \ref{FH10} and \ref{FH15}) longitudinal intervals.   

The degree of linear polarization is considerably higher than that of circular polarization for most of these examples.  There are however cases such as that depicted in Fig.~\ref{FH10} for which the Stokes parameter $V$ is comparable to, or greater than, the Stokes parameter $L$ over parts of the pulse window.  As can be seen from Figs.~\ref{FH7}--\ref{FH10} and \ref{FH13}--\ref{FH15}, the sense of circular polarization is often reversed across certain components of the pulse profile.   

S-shaped swings through $180^\circ$ are frequently exhibited by the longitudinal distributions of the polarization position angles of either one (Figs.~\ref{FH7} and  \ref{FH8}) or both (Figs.~\ref{FH10}--\ref{FH14}) polarization modes.  But there are also cases in which the position angles of either one or both modes remain constant across the pulse profile (Figs.~\ref{FH7} and~\ref{FH9}).  Position angles of the two concurrent modes can be approximately orthogonal over a section of the pulse profile (Figs.~\ref{FH7}, \ref{FH8}, \ref{FH11}, \ref{FH13} and \ref{FH14}) or can be coincident (Figs.~\ref{FH9} and \ref{FH12}).  Note that these properties of the polarization position angle stem from the interplay between the stationary points of various phase functions rather than from the orientation of any magnetic field.

Though occurring in every case, the rapid small-amplitude modulations of the pulse profiles and position-angle distributions are only visible in Figs.~\ref{FH7}--\ref{FH9} and \ref{FH11} for which $\kappa_{\rm u}\le10^4$.  The wavelength of such modulations is proportional to $\kappa_{\rm u}^{-1}$, so that the microstructure they superpose on pulse profiles and position-angle distributions is too sharp and dense to show up in the rest of the plotted figures.  The free parameter $\kappa_{\rm u}^{-1}$ cannot of course be smaller than the (unknown) thickness of the current sheet in units of the light-cylinder radius. 

It should be added that at any given value of the inclination angle $\alpha$, the pulse observed at $\pi-\theta_P$ differs from that observed at $\theta_P$ only in that the intensity $V$ of its circularly polarized part is replaced by $-V$ and its longitude $\varphi_P$ is replaced by $\varphi_P+\pi$.  Moreover, the results for $\alpha > \pi/2$ follow from those for $\alpha < \pi/2$ by replacing $\theta_P$, $\varphi_P$ and $V$ by  $\pi-\theta_P$, $\varphi_P+\pi$ and $-V$, respectively. 

Figures~\ref{FH15} and~\ref{FH16} illustrate an example of a radically different type of pulse: one detectable near those observation points for which two nearby stationary points of the phase of one of the oscillating exponentials described in Section~\ref{sec:volume} coalesce, thus giving rise to a much tighter focusing of the emitted waves.  For any given values of the inclination angle $\alpha$ and the observer's distance $R_P$, there are at least two, and at most eight, critical values ($\theta_{PS}$) of the colatitude $\theta_P$ of the observation point at which two nearby stationary points of one of the phase functions in question coalesce to give rise to a higher-order focusing of this kind.  Though their profiles over the entire pulse window look similar to those of other pulses (Fig.~\ref{FH15}), such pulses display extraordinarily large amplitudes and short widths once they are inspected over sufficiently short longitudinal intervals to resolve their peaks (Fig.~\ref{FH16}a).  We shall see below that the extraordinary values of the amplitudes and widths of such pulses, illustrated by the example in Figs.~\ref{FH15} and~\ref{FH16}a, are what underpin the high brightness temperatures and broad frequency spectra of the radiation generated by the current sheet. 

\begin{figure*}
\centerline{\includegraphics[width=11cm]{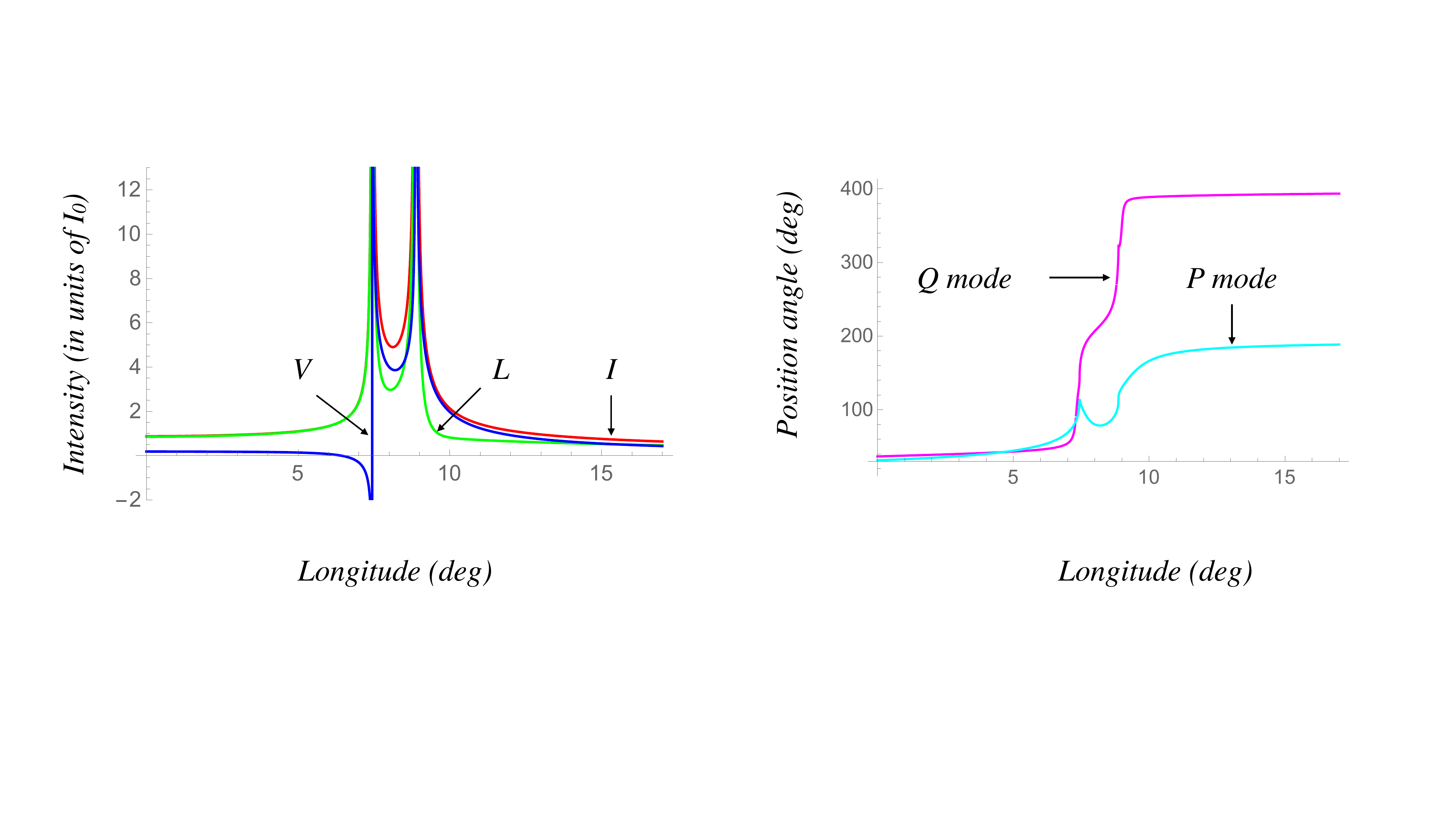}}
\caption{The Stokes parameters $(I, V, L)$ and the position angles $\psi$ of the polarization modes $P$ and $Q$ at an observation point with the colatitude $\theta_P=55^\circ$ for the inclination angle $\alpha=45^\circ$ and $\kappa_{\rm u}=10^2$.  Note that the polarization of this pulse changes from linear to circular across it and the position angle of the Q mode swings through $180^\circ$ across each one of its two components.  The pulse window encompasses another pulse at a longitudinal distance of about $115^\circ$ from this one}
\label{FH10}
\end{figure*} 

\begin{figure*}
\centerline{\includegraphics[width=11cm]{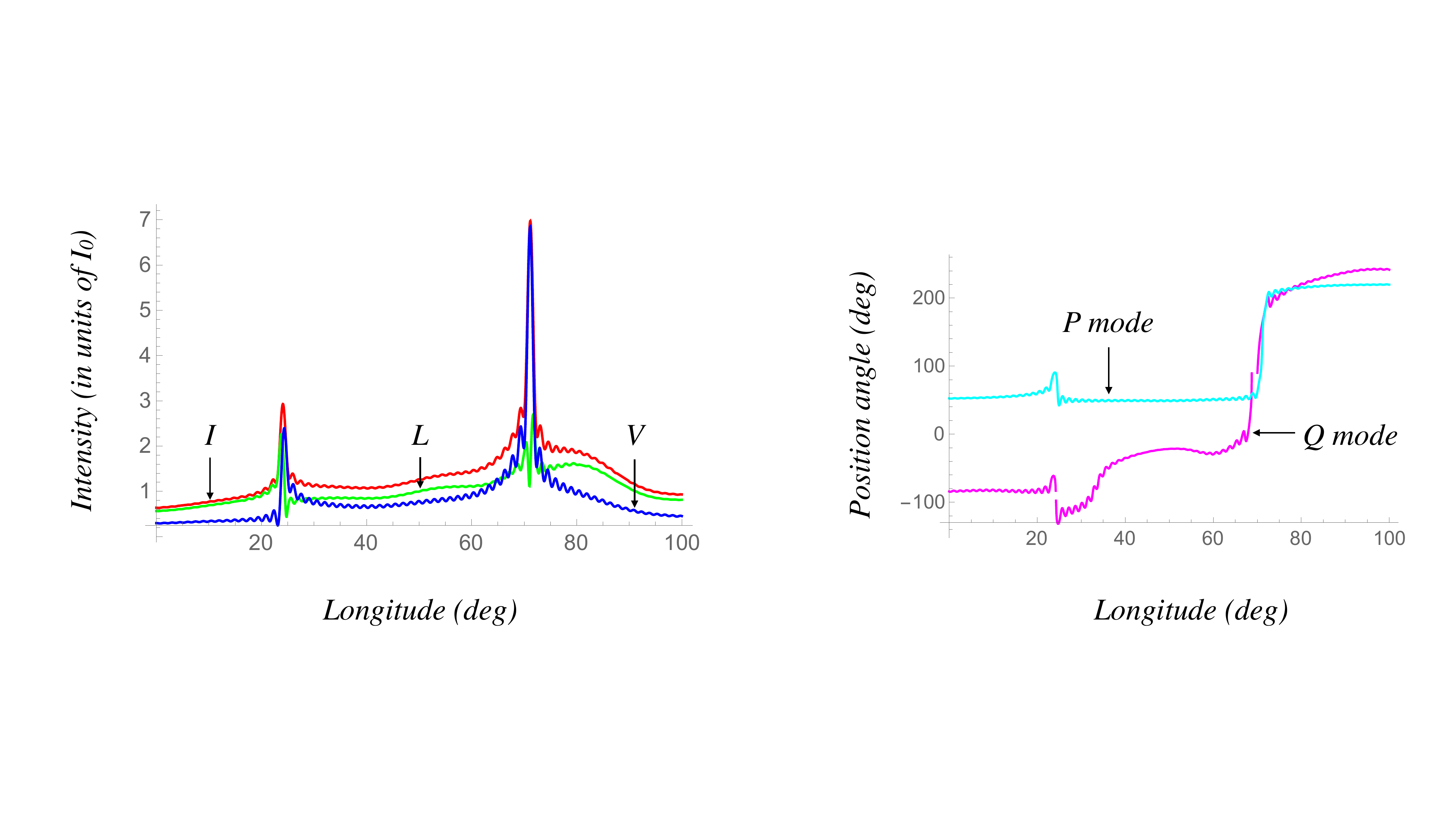}}
\caption{The Stokes parameters $(I, V, L)$ and the position angles $\psi$ of the polarization modes $P$ and $Q$ at an observation point with the colatitude $\theta_P=35^\circ$ for the inclination angle $\alpha=45^\circ$ and $\kappa_{\rm u}=10^2$.  Note the high degrees of circular polarization of both components of this pulse.  Position angle of the $Q$ mode turns through $360^\circ$ across the narrow gap at longitude $70^\circ$ in the curve depicting its distribution.  These are no other components in the pulse window.}
\label{FH11}
\end{figure*}

\subsection{Brightness temperature}
\label{subsec:brightness}

The brightness temperature $T_{\rm b}$ of the present radiation can be accurately determined by equating the magnitude of the Poynting flux of this radiation to the Rayleigh-Jeans law for the energy that a black body of the same temperature would emit per unit time per unit area into the frequency band $\Delta\nu$ centred on the frequency $\nu$.  The resulting equation in conjunction with Eq.~(\ref{E1}) shows that $T_{\rm b}$ is related to the dimensionless Stokes parameter ${\hat I}=I/I_0$ by
\begin{equation}
T_{\rm b}=\frac{c\,\omega^2 r_{s0}^4B_0^2}{8\pi k_{\rm B}{R}_P^2\nu^2\Delta \nu}{\hat I},
\label{E2}
\end{equation}
where ${r}_{s0}$, $B_0$, $R_P$ and $k_{\rm B}$ are radius of the star, magnitude of the star's dipolar field at its magnetic pole, distance of the observer from the source and the Boltzmann constant, respectively.  

Replacing the parameters that appear in Eq.~(\ref{E2}) by
\begin{equation}
B_0=10^{12}{\hat B}_0\,\,{\rm Gauss},\quad \omega=10^2{\hat P}^{-1}\,\,{\rm rad}/{\rm s},
\label{E3}
\end{equation}
\begin{equation}
R_P=D\,\,{\rm kpc}=3.085\times10^{21}D\,\,{\rm cm},
\label{E4}
\end{equation}
\begin{equation}
r_{s0}=10^6 d\,\,{\rm cm},\,\,\nu=10^8 {\hat\nu}\,\,{\rm Hz},\,\, \Delta\nu=10^7\Delta{\hat\nu}\,\,{\rm Hz},
\label{E5}
\end{equation}
we obtain 
\begin{equation}
T_{\rm b}=9.11\times10^{10}\,{\hat I}\,{\hat T}_{\rm b}
\label{E5p5}
\end{equation}
with
\begin{equation}
{\hat T}_{\rm b}=\frac{{\hat B}_0^2d^4}{{\hat P}^2D^2{\hat \nu}^2\Delta{\hat\nu}},
\label{E5p9}
\end{equation}
in which the value of ${\hat I}$ is specified, as in the case of Figs.~\ref{FH7}--\ref{FH15}, by the numerical evaluation of the Stokes parameter $I$ in units of $I_0$ at the highest resolved peak of the pulse detected at ${\hat R}_P=R_P\omega/c=10^{13}$ (i.e., at ${\hat R}_P=1.028\times10^{13}D{\hat P}^{-1}$ when the factor $1.028D{\hat P}^{-1}$ equals unity). 

The brightness temperatures implied by Eq.~(\ref{E5p5}) and $\kappa_{\rm u}=10^7$ are listed in Table~\ref{T1} for the pulses depicted in Figs.~\ref{FH7}, \ref{FH10} and \ref{FH15} and for a pair of examples of the pulses that are detected at the critical colatitudes $\lim_{R_P\to\infty}\theta_{PS}$.  Table~\ref{T1} also shows the full width at half maximum $\delta\varphi_P$ of the components of the listed pulses that have the highest resolved peaks (see Fig.~\ref{FH16}).  Note that, as indicated by the last column of Table~\ref{T1}, the pulse profiles depicted in Figs.~\ref{FH7}--\ref{FH15} have to be plotted on considerably shorter longitudinal scales before their peaks assume the shapes shown in Fig.~\ref{FH16} and the maximum values of their dimensionless intensity ${\hat I}$ can be discerned graphically.  In general, as one reduces the longitudinal interval over which ${\hat I}$ is plotted, the peak of the pulse splits in two before the finite widths of either of the partitioned pulses are visible.  Once resolved, the longitudinal distributions of the narrow pulses that are observed at critical colatitudes all have the same shape as that of the pulse shown in Fig.~\ref{FH16}a.  

\begin{table*}
\centering
\caption{Brightness temperature $T_{\rm b}$ and full width at half maximum $\delta\varphi_P$ of the pulse detected at colatitude $\theta_P$ for the inclination angle $\alpha$ and ${\hat R}_P=10^{13}$, $\kappa_{\rm u}=10^7$.  The dimensionless factor ${\hat T}_{\rm b}$, defined by Eqs.~(\ref{E2})--(\ref{E5p9}), is of the order of unity in the case of most radio pulsars.  The limiting values of the critical colatitudes for the values $30^\circ$ and $75^\circ$ of $\alpha$ in the second column are $33.932818533330613261^\circ$ and $48.533945294618400228^\circ$, respectively.}
\label{T1}
\begin{tabular}{ c | c | c | c }  
\hline   
$\alpha\,\, ({\rm deg})$ & $\theta_P\,\, ({\rm deg})$ & $T_{\rm b}/{\hat T}_{\rm b}\,\, ({}^{\circ}{\rm K})$ & $\delta\varphi_P\,\, ({\rm deg})$ \\       \hline
 $5$ & $5$ & $9.56\times10^{22}$ & $ 6.76\times10^{-9} $\\  \hline
 $30$ & $\lim_{R_P\to\infty}\theta_{PS}$ & $5.1\times10^{41}$ & $ 2.39\times10^{-24} $\\  \hline
 $45$ & $55$ & $1.73\times10^{18}$ & $ 3.81\times10^{-7} $ \\  \hline
 $60$ & $90$ & $8.67\times10^{40}$ & $ 6.94\times10^{-25} $ \\  \hline
 $75$ & $\lim_{R_P\to\infty}\theta_{PS}$ & $3.01\times10^{39}$ & $ 5.73\times10^{-25} $ \\  \hline
\end{tabular}
\end{table*}

It should be noted that evaluating the critical colatitude $\theta_{PS}(\alpha,{\hat R}_P)$ at ${\hat R}_P=\infty$ is not the same as placing the observation point at infinity.  For any given value of $\alpha$, as one moves away from the source along a critical colatitude that is evaluated at ${\hat R}_P=\infty$, the separation between the nearby stationary points of the relevant phase function continually decreases with distance until it reduces to zero at infinity.  In every case, the separation between the stationary points in question has a very small value at all distances ${\hat R}_P\gg1$ along the latitudinal direction $\lim_{R_P\to\infty}\theta_{PS}$.  Since the coherence of the present radiation stems from the closeness of two stationary points of a phase function, the brightness temperature $T_{\rm b}$ in Eq.~(\ref{E5p5}) thus has a high value in the latitudinal direction $\lim_{R_P\to\infty}\theta_{PS}$ at all observation points in the far zone.

\begin{figure*}
\centerline{\includegraphics[width=11cm]{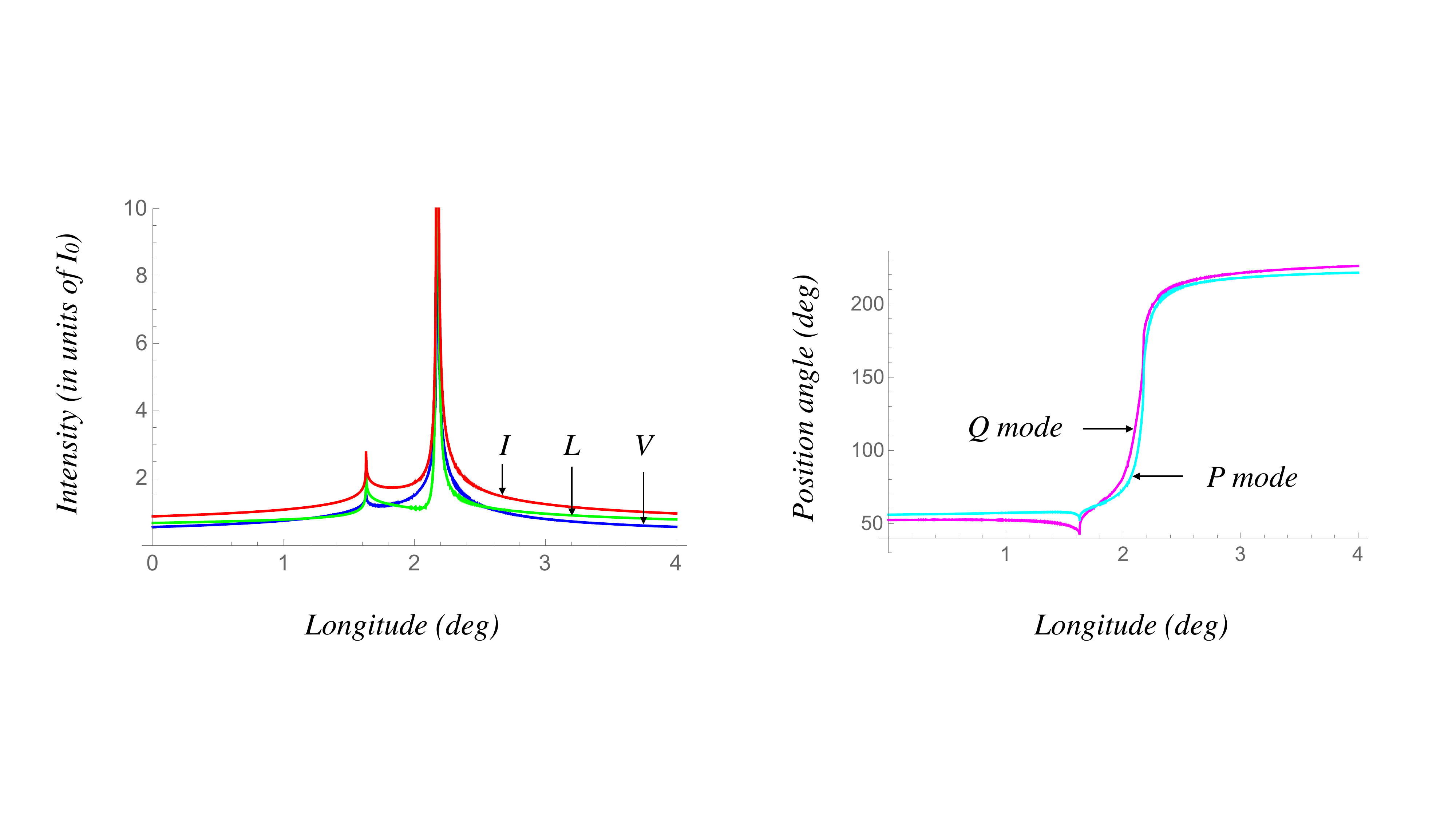}}
\caption{The Stokes parameters $(I, V, L)$ and the position angles $\psi$ of the polarization modes $P$ and $Q$ at an observation point with the colatitude $\theta_P=30^\circ$ for the inclination angle $\alpha=65^\circ$ and $\kappa_{\rm u}=10^2$.  Note the simultaneous swing through $180^\circ$ of the position angles of both modes across the right-hand component of the pulse.  Note also the high degree of circular polarization of the pulse throughout the depicted longitudes.  The pulse window encompasses in addition a weaker pulse at a longitudinal distance of about $20^\circ$ from these ones.}
\label{FH12}
\end{figure*} 

\begin{figure*}
\centerline{\includegraphics[width=11cm]{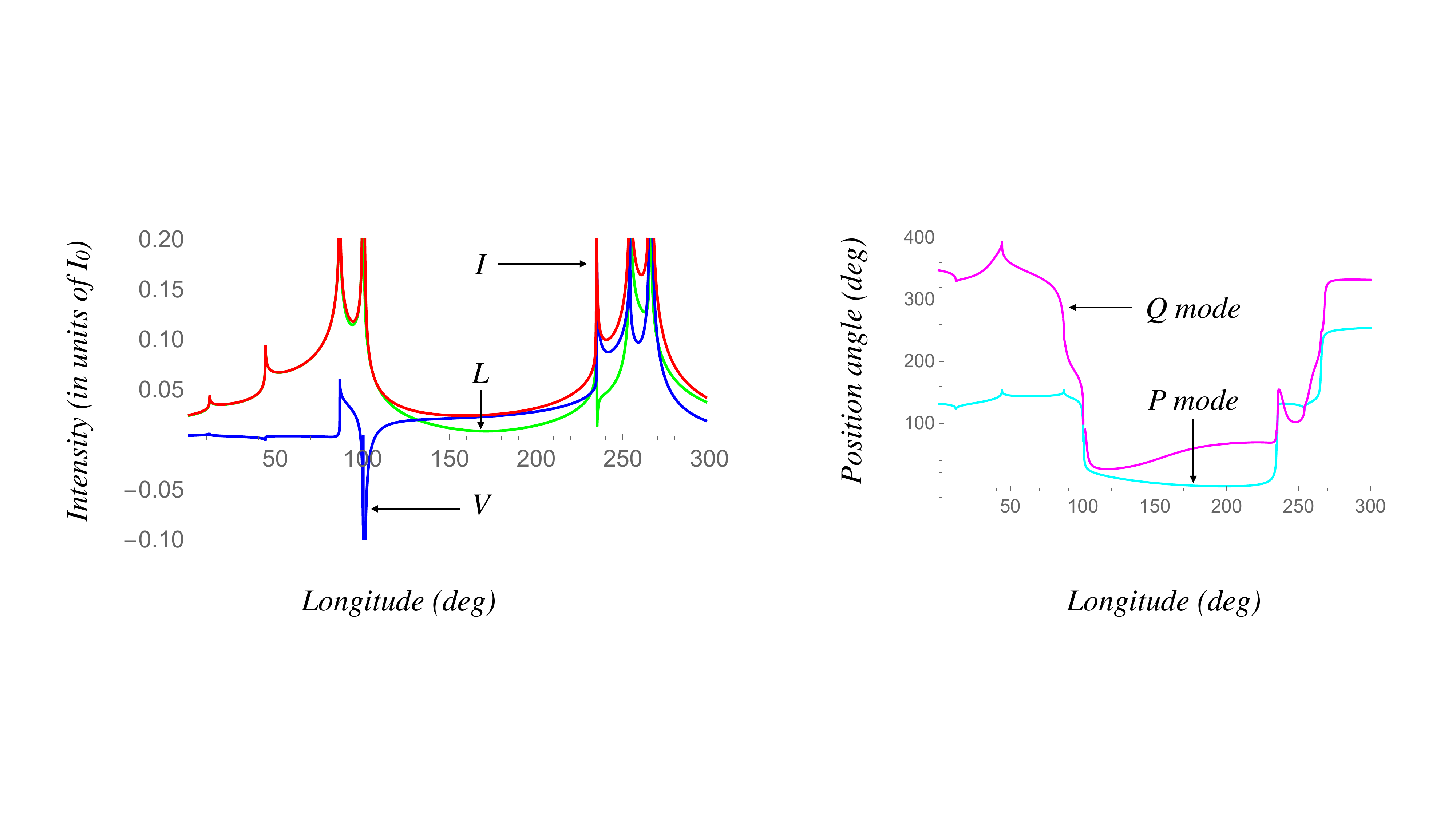}}
\caption{The Stokes parameters $(I, V, L)$ and the position angles $\psi$ of the polarization modes $P$ and $Q$ at an observation point with the colatitude $\theta_P=77.5^\circ$ for the inclination angle $\alpha=65^\circ$ and $\kappa_{\rm u}=10^7$.  In this example of a multi-peak pulse profile the stationary points of all of the phase functions described in Section~\ref{sec:volume} contribute toward the field.  Position angle of the $Q$ mode turns through $360^\circ$ across the narrow gap at longitude $100^\circ$ in the curve depicting its distribution.}
\label{FH13}
\end{figure*} 

\subsection{Frequency spectrum}
\label{subec:frequency}

Given that the radiation field of the current sheet depends on the observation time $t_P$ and the azimuthal coordinate $\varphi_P$ of the observation point only in the combination $\varphi_P-\omega t_P$, the frequency spectrum of this radiation is equally well described by the Fourier decomposition of its longitudinal distribution.   Depending on the colatitude of the observation point, the content of this spectrum stems either from the fluctuations manifested in the microstructure of the pulse profile or from the much sharper variations characterised by the full width at half maximum ($\delta\varphi_P$) of the pulse components that arise from isolated or imminently coalescing stationary points of a phase function (Table~\ref{T1}).  

For frequencies much higher than the rotation frequency $\omega/2\pi$, the spectral flux density $S_\nu$ of the radiation has the power-law dependence $\nu^{-\beta}$ on the frequency $\nu$ with a spectral index $\beta$ that assumes the following values in various ranges of frequency: $2/3$, $1$, $4/3$, $5/3$, $2$, $7/3$ (see~\cite{Ardavan2021}, Section~5.3).  

At colatitudes close to the critical angles $\theta_{PS}$ where two stationary points of a phase function are nearly coincident, the narrow widths of the peaks of the time-domain pulse profiles (Figs.~\ref{FH16} and~\ref{FH18}) are reflected in a broad frequency spectrum that extends from radio waves to gamma rays.  This result, not only follows from the expression for the spectrum whose derivation was outlined in Section~\ref{sec:volume}, but is also illustrated by the example shown in Fig.~\ref{FH16}a: the width $\delta\varphi_P=1.21\times10^{-26}$~rad of the pulse depicted in that figure implies a frequency spectrum that extends as far as  $\nu\simeq\omega/(2\pi\delta\varphi_P)\simeq1.31\times10^{27}$ Hz when $\omega=10^2$ rad/s. 

Figure~\ref{FH17} shows the on-pulse spectral distribution of the radiation that is detectable close to a critical colatitude $\theta_{PS}$ over a high-frequency band.  The peak amplitudes of the oscillations of this spectrum, which decrease as $\nu^{-2/3}$ with increasing frequency $\nu$, are manifested in a dynamic spectrum as a set of discrete frequency bands (\cite{Ardavan2021}, Section 5.3).    

Figures~\ref{Crab}, \ref{Radio1}, \ref{Radio2} and \ref{Radio3} show examples of fitting the observed data on gamma-ray and radio spectra of pulsars with the spectral distribution function of the present radiation given in Eq.~(177) of~\cite{Ardavan2021}.

\subsection{A phase lag between low-frequency and high-frequency peaks of the pulse profile}
\label{subsec:phase-lag}

Given that the sensitivity of any detector with which they are observed is limited to a finite bandwidth, the time-domain pulse profiles depicted in Figs.~\ref{FH7}--\ref{FH15} do not have the same shapes in the frequency domain.  The range of frequencies over which a given component of a time-domain pulse profile is observable depends on the full width at half maximum of the peak of that pulse component: a width illustrated in Fig.~\ref{FH16} that can be determined by  plotting the distribution of the Stokes parameter $I$ over successively shorter longitudinal intervals centred on the peak of the distribution until the maximum value of $I$ stops growing.

The full width at half maximum of the two higher peaks of the pulse depicted in Fig.~\ref{FH15} (those at longitudes $30^\circ$ and $210^\circ$) is considerably shorter than that of the two lower peaks of this pulse (at longitudes $15^\circ$ and $195^\circ$):  while $\delta\varphi_P=1.19\times10^{-29}$ rad for the higher peaks, the corresponding width of the lower peaks is $\delta\varphi_P=6\times10^{-8}$ rad when $\kappa_{\rm u}=10^{10}$.  Thus the two higher peaks of this time-domain pulse which are detectable in the gamma-ray band (at frequencies around $\omega/(2\pi\delta\varphi_P)=1.33\times10^{30}{\hat P}^{-1}$ Hz) lag its two lower peaks which are detectable in the radio band (at frequencies around $2.65\times10^8{\hat P}^{-1}$ Hz) by $15^\circ$.

Figure~\ref{FH18} (in which ${\hat R}_P=10^{13}$, $\theta_P=\lim_{R_P\to\infty}\theta_{PS}$, $\alpha=50^\circ$, $\kappa_{\rm u}=10^{10}$) shows another example of such a phase lag, one in which the radio pulse is double-peaked but the gamma-ray pulse has a single component.  In Fig.~\ref{FH18} we have plotted the contributions toward the Stokes parameter $I$ that arise from two isolated stationary points of one of the phase functions in part (a) and those from two nearly coalescent stationary points of another phase function in part (b).  Once resolved, the two radio peaks at longitudes $30^\circ$ and $95^\circ$ (in Fig.~\ref{FH18}a) each have the width $\delta\varphi_P=3.36\times10^{-9}$ rad while the single gamma-ray peak at longitude $112^\circ$ of this profile (in Fig.~\ref{FH18}b) has the width $\delta\varphi_P=1.83\times10^{-29}$ rad.  The longitudes of the adjacent radio and gamma-ray peaks are thus separated by $17^\circ$ in this case (\cite{Ardavan2021}, Section 5.4). 

\begin{figure*}
\centerline{\includegraphics[width=11cm]{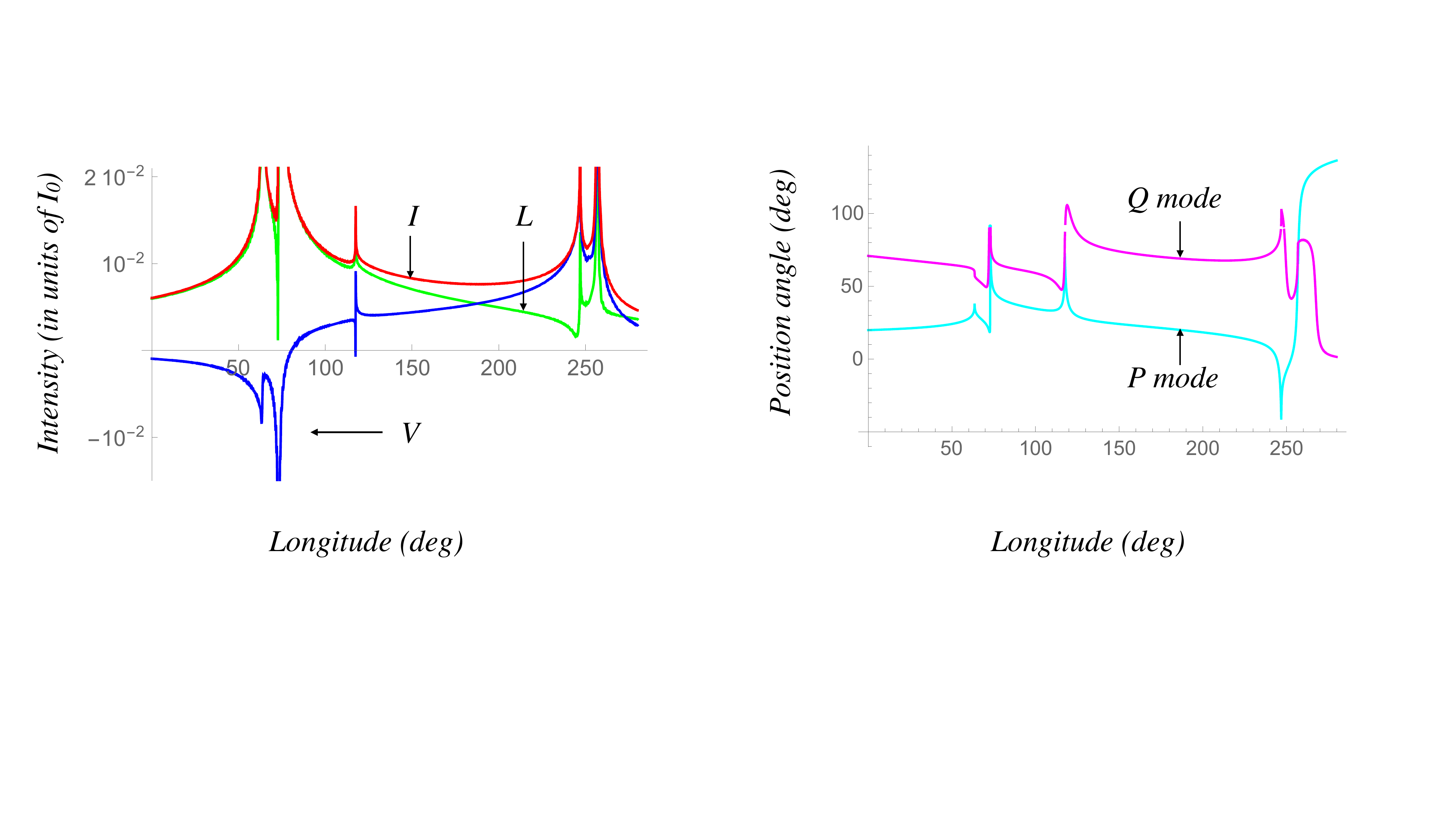}}
\caption{The Stokes parameters $(I, V, L)$ and the position angles $\psi$ of the polarization modes $P$ and $Q$ at an observation point with the colatitude $\theta_P=100^\circ$ for the inclination angle $\alpha=80^\circ$, $\kappa_{\rm u}=10^4$ (pulse profile) and $\kappa_{\rm u}=10^7$ (position angle).  In this case, the stationary points of all of the phase functions described in Section~\ref{sec:volume} contribute toward the field.  This is an example of a multi-component pulse the polarization position angles of whose different components have differing longitudinal variations.}
\label{FH14}
\end{figure*} 

\begin{figure*}
\centerline{\includegraphics[width=11cm]{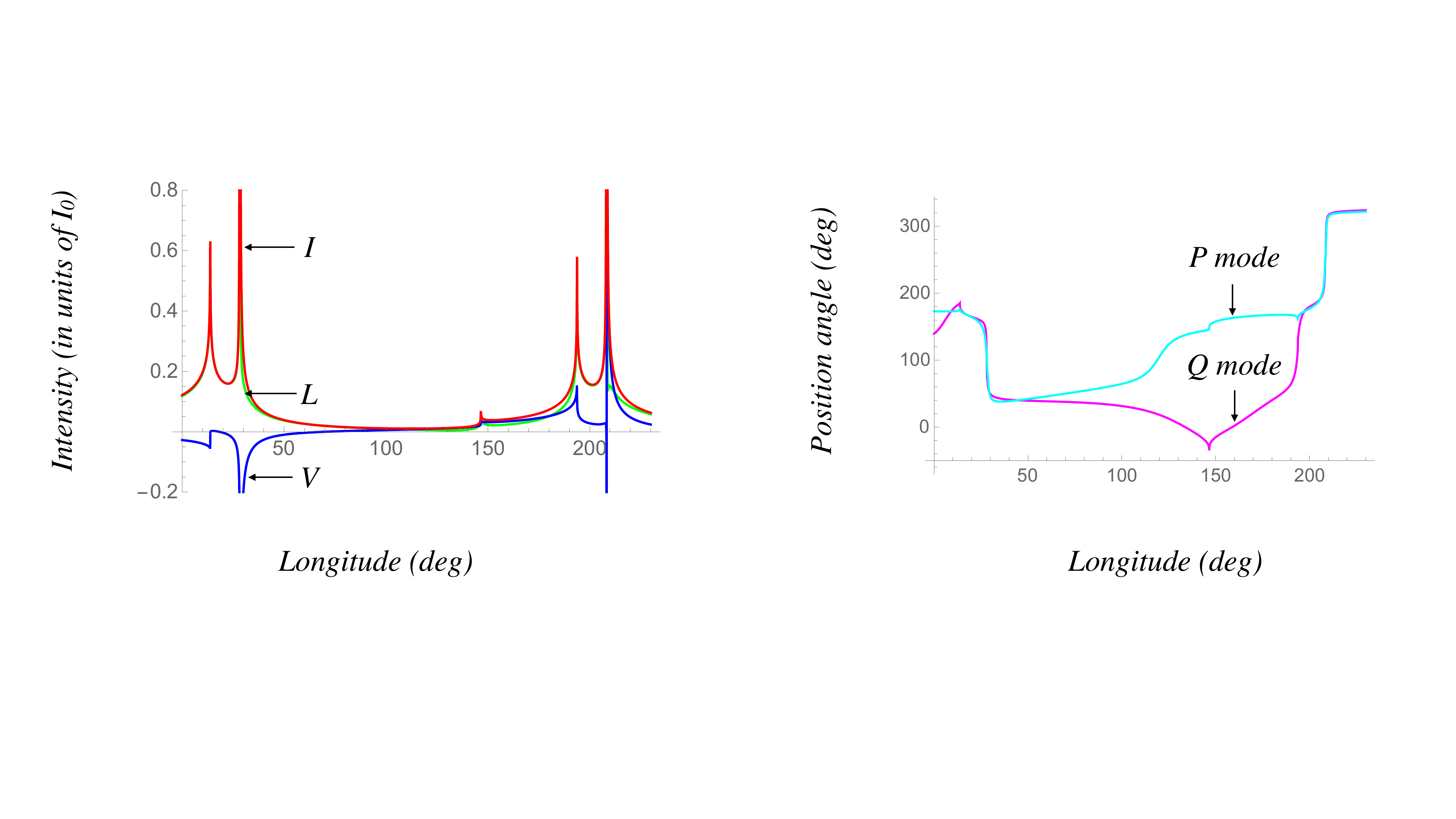}}
\caption{The Stokes parameters $(I, V, L)$ and the position angles $\psi$ of the polarization modes $P$ and $Q$ at an observation point with the coordinates ${\hat R}_P=10^{13}$ and $\theta_P=90^\circ$ for the inclination angle $\alpha=60^\circ$ and $\kappa_{\rm u}=10^7$.  In this case, the stationary points of all of the phase functions described in Section~\ref{sec:volume} contribute toward the field.  The two higher peaks of this profile (at longitudes $30^\circ$ and $210^\circ$) arise from nearly coalescent stationary points of one of these phase functions.  The lower peaks (at longitudes $15^\circ$ and $195^\circ$), on the other hand, arise from two isolated stationary points of another phase function.  The full widths at half maxima of the higher and lower peaks of this pulse profile can be inferred from Figs.~\ref{FH16}a and \ref{FH16}b, respectively.}
\label{FH15}
\end{figure*} 

\begin{figure*}
\centerline{\includegraphics[width=13cm]{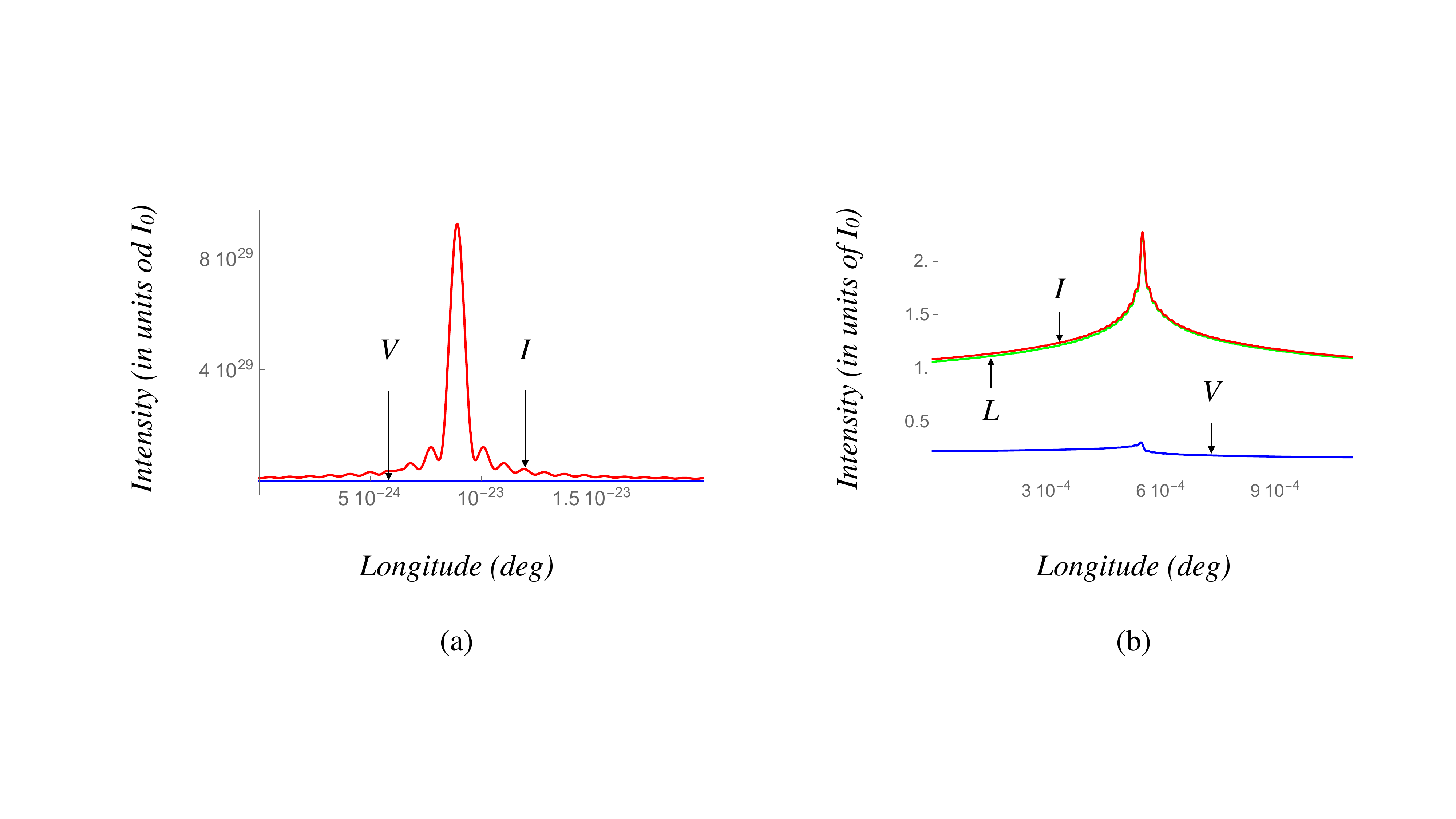}}
\caption{In part (a) of this figure, the width of the component at longitude $210^\circ$ of the distribution of the Stokes parameter $I$ that is depicted in Fig.~\ref{FH15} is resolved by plotting this distribution over successively shorter longitudinal intervals centred at its peak until the value of $I$ at its maximum stops growing.  In part (b), the same procedure is applied to the component at longitude $195^\circ$ of the distribution of the Stokes parameter $I$ depicted in Fig.~\ref{FH15}.  Besides the difference in width, there is a generic difference between the shape of a resolved pulse that arises from two nearly coalescent stationary points of a phase function, as in part (a), and that of a resolved pulse that arises from an isolated stationary point of a phase function, as in part (b).}
\label{FH16}
\end{figure*}

\subsection{Flux density and its rate of decay with distance}
\label{subsec:inverse-square}

Flux density of a radiation is given by the magnitude of the Poynting vector.  In the present case, however, the linear extents in the azimuthal direction, $R_P\delta\varphi_P$, of the focused radiation beams that embody the high-frequency (optical to gamma-ray) radiation are invariably smaller than $1$ cm at $R_P=1$~kpc (see Table~\ref{T1}).  The amount of energy that crosses a unit area per unit time is therefore given by 
\begin{eqnarray}
S&=&\frac{c}{4\pi}\vert{\bf E}^{\rm uc}\vert^2\delta\varphi_P\nonumber\\*
&=&2.79\times10^{-3}{\hat I}\delta\varphi_P{\hat S} \quad {\rm erg \,cm^{-2} s^{-1}}
\label{E6}
\end{eqnarray}
in which ${\bf E}^{\rm uc}$ is the electric-field vector of the unconventional component of the radiation,
\begin{equation}
{\hat S}=\left(\frac{{\hat B}_0d^2}{{\hat P}D}\right)^2,
\label{E7}
\end{equation}
${\hat I}=I/I_0$ and $\delta\varphi_P$ is in radians [see Eqs. (\ref{E1}) and (\ref{E3})--(\ref{E5p9})].  

Note that the linear extents in the latitudinal direction, $R_P\delta\theta_P$, of the focused radiation beams that embody the high-frequency radiation are of the order of the light-cylinder radius, $c/\omega$, in general: these beams remain fully in focus at all distances ${\hat R}_P$ within a latitudinal interval $\delta\theta_P\simeq\vert\theta_P-\theta_{PS}\vert$ which turns out to be of the order of ${\hat R}_P^{-1}$ independently of the values of the other parameters (see Fig.~\ref{FH6}b).  In the case of $\alpha=60^\circ$, $\theta_P=90^\circ$, $\kappa_{\rm u}=10^7$ and $D=1$ kpc depicted in Figs.~\ref{FH15} and~\ref{FH16}, for example, the flux density $S$ has the value $32.1\, {\hat S}$ erg cm${}^2$ s${}^{-1}$.  At latitudes closer to, or further away from, the critical angle for this example ($\lim_{R_P\to\infty}\theta_{PS}=90^\circ$), the degree of focusing of the radiation beam and so the value of the flux density $S$ is, respectively, higher or lower.

As pointed out in Section~\ref{sec:volume}, the length of the interval $\vert\theta_{{\rm max}}-\theta_{{\rm min}}\vert$ separating the $\theta$ coordinates of the maximum and minimum of a phase function decreases as ${\hat R}_P^{-1/2}$ with increasing ${\hat R}_P$ when this interval is small, i.e., when the colatitude $\theta_P$ of the observation point lies within an interval of length ${\hat R}_P^{-1}$ of the critical angle $\theta_{PS}(\alpha,L)$ with $L > {\hat R}_P$ (see Fig.~\ref{FH6}).  In particular, if the observation point has the colatitude $\lim_{R_P\to\infty}\theta_{PS}$ (or $\pi-\lim_{R_P\to\infty}\theta_{PS}$), then the maximum and minimum of the phase function in question coalesce into an inflection point only at ${\hat R}_P\to\infty$, rather than at a finite distance $L$.  In the case illustrated in Figs.~\ref{FH15} and~\ref{FH16}, for example, the colatitude of the observation point equals $\lim_{R_P\to\infty}\theta_{PS}=90^\circ$, so that at the finite distance ${\hat R}_P=10^{13}$ the $\theta$ coordinates of  maximum and minimum of the phase are separated by the short interval $3.05\times10^{-5}$~deg. In fact, this separation has the value $3.05\times10^{-5}({\hat R}_P/10^{13})^{-1/2}$~deg for all ${\hat R}_P$.

\begin{figure}
\centerline{\includegraphics[width=10cm]{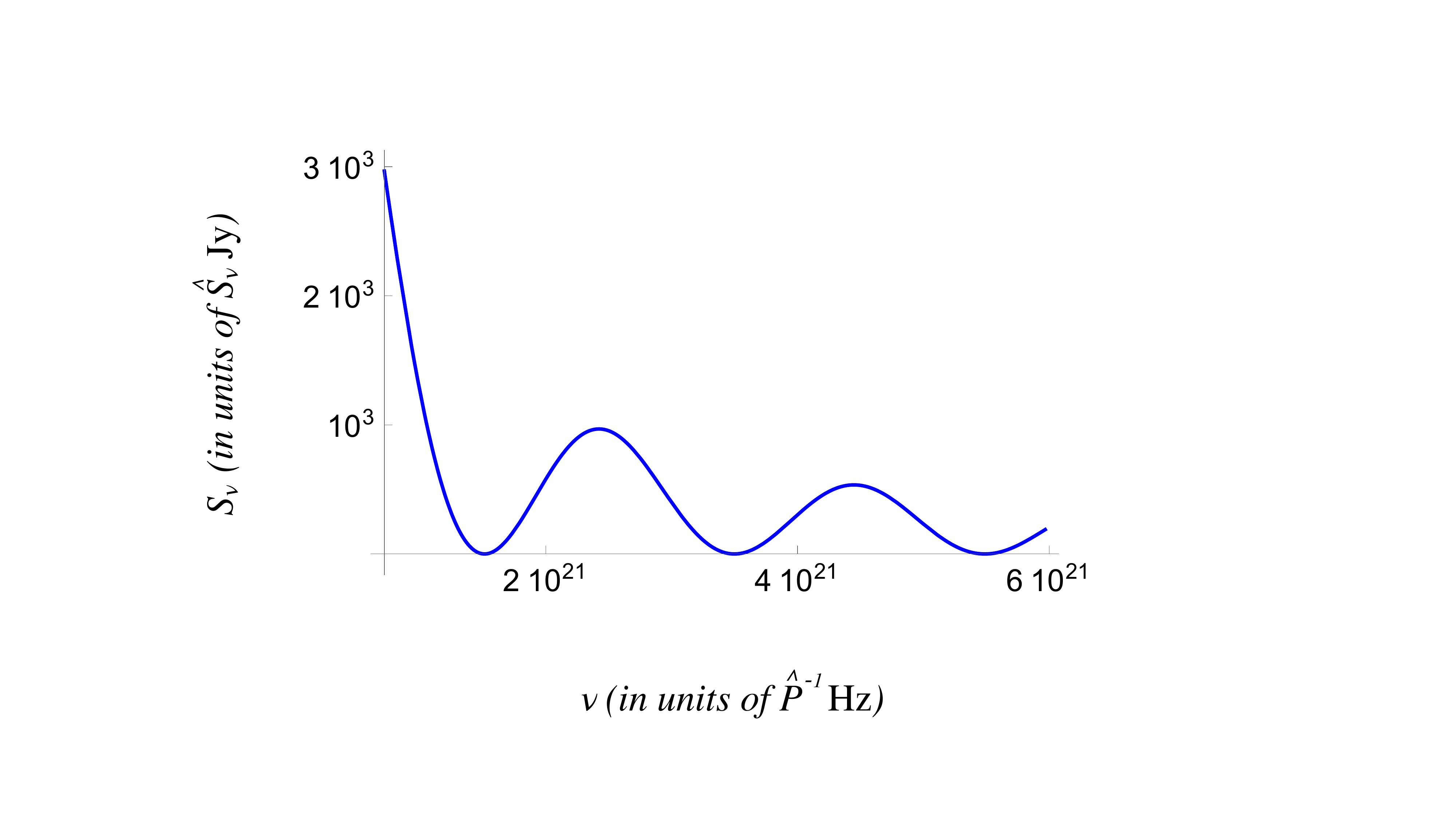}}
\caption{Distribution of the spectral flux density of the narrow pulse shown in Fig.~\ref{FH16}a.  Thus the dynamic spectrum of the radiation consists of a discrete set of bands whose amplitudes decrease with increasing frequency.}
\label{FH17}
\end{figure} 

\begin{figure}
\centerline{\includegraphics[width=10cm]{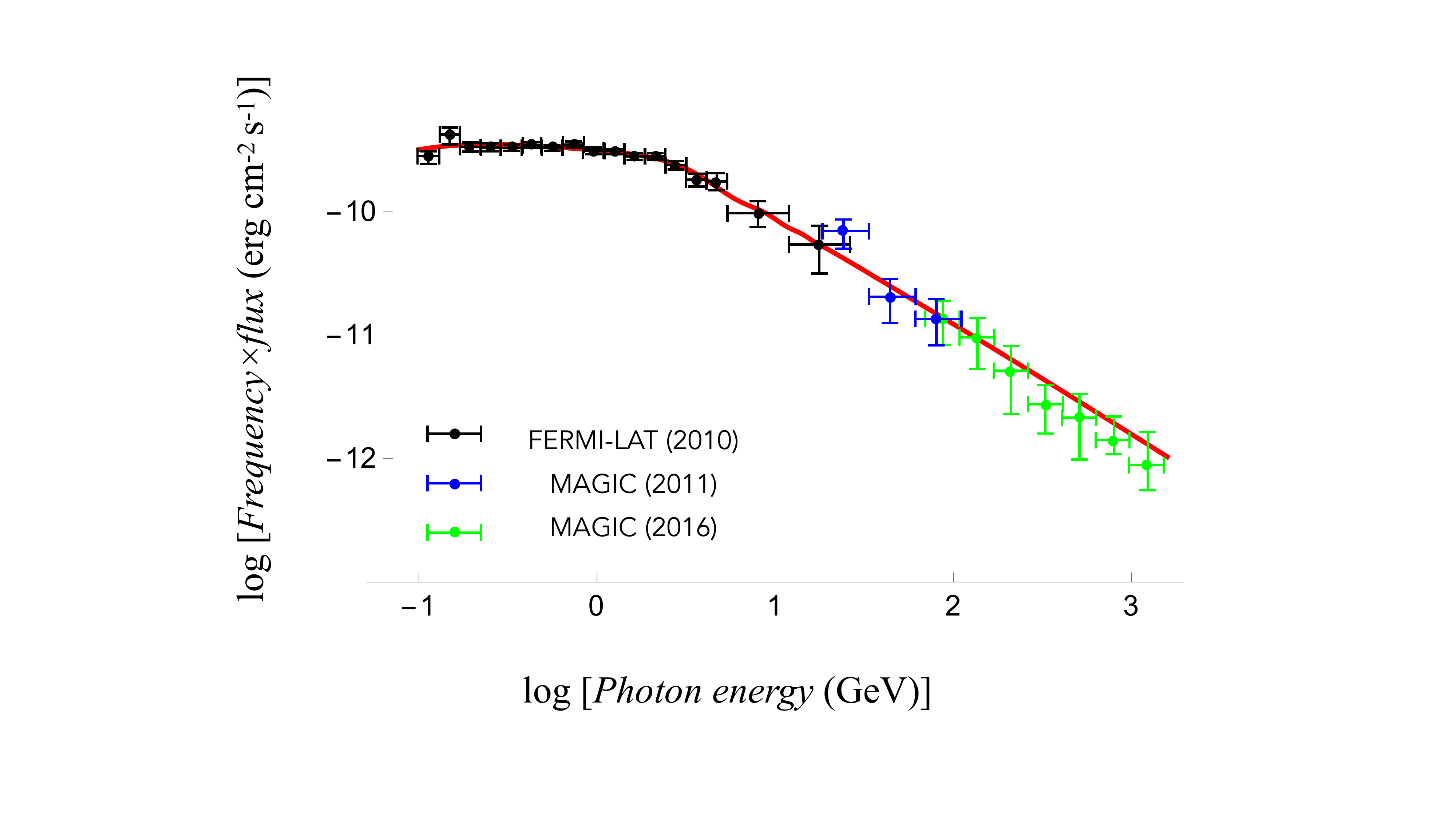}}
\caption{Gamma-ray spectrum of the Crab pulsar fitted with the spectral energy distribution of the emission from its current sheet (see~\cite{ArdavanCVG} for corresponding spectra of the Vela and Geminga pulsars and for further details).}
\label{Crab}
\end{figure} 

\begin{figure}
\centerline{\includegraphics[width=9cm]{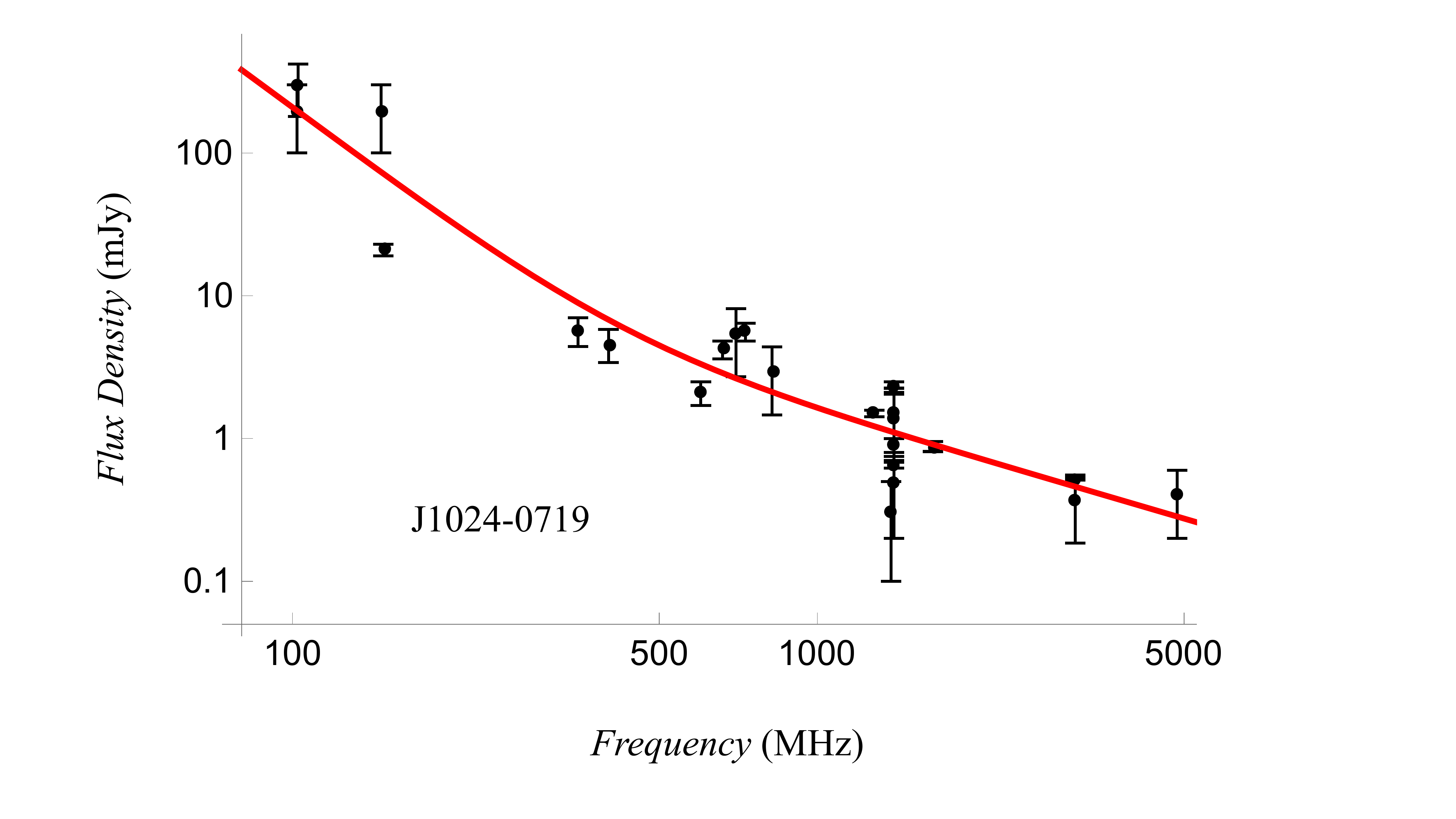}}
\caption{Radio spectrum of PSR J1024-0719 fitted with the spectral distribution function of the emission from its current sheet (see~\cite{Ardavan2024Radio} for other fitted radio spectra and for further details).}
\label{Radio1}
\end{figure} 

\begin{figure}
\centerline{\includegraphics[width=9cm]{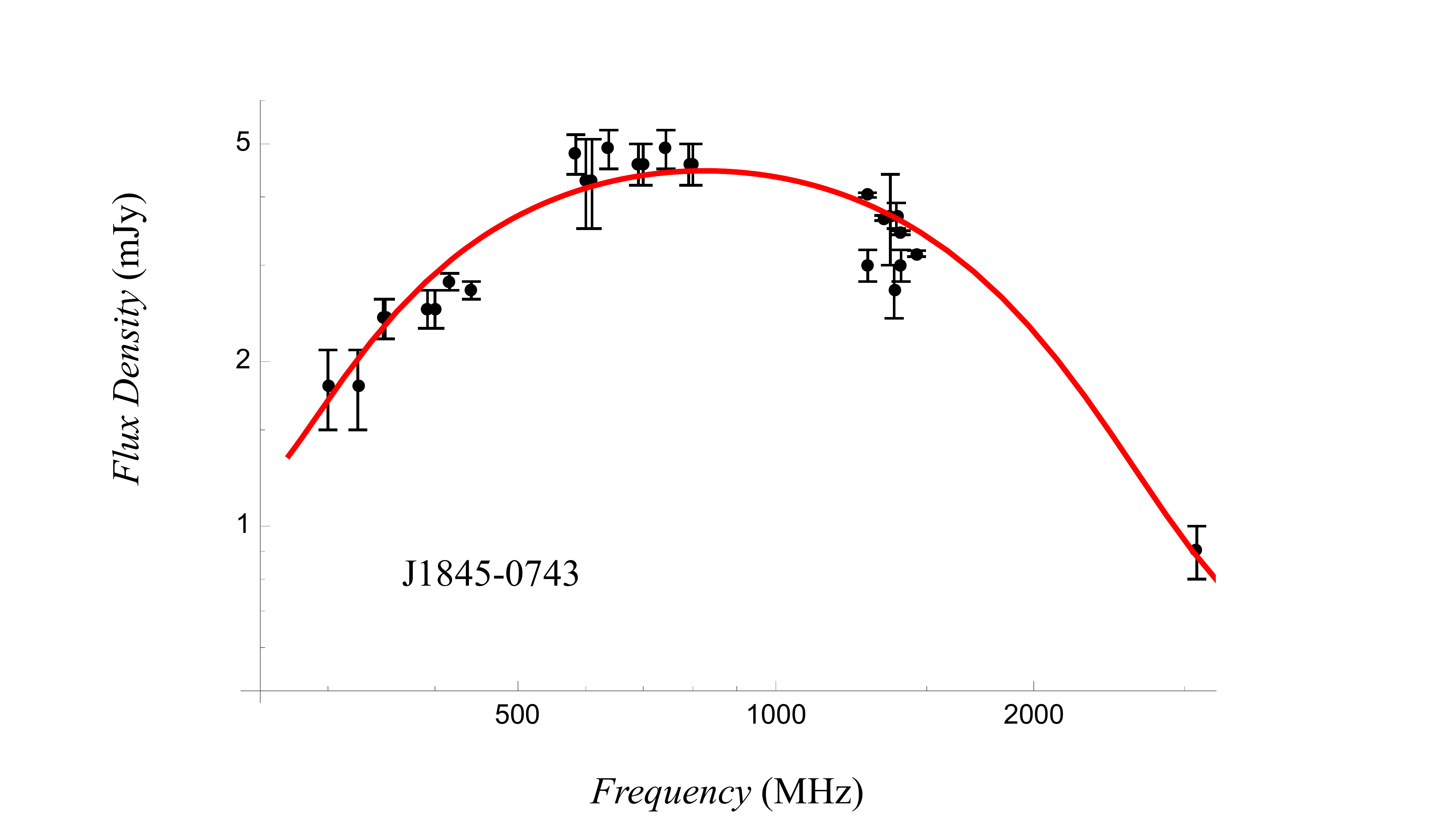}}
\caption{Radio spectrum of PSR J1845-0743 fitted with the spectral distribution function of the emission from its current sheet (see~\cite{Ardavan2024Radio} for other fitted radio spectra and for further details).}
\label{Radio2}
\end{figure} 

\begin{figure}
\centerline{\includegraphics[width=8cm]{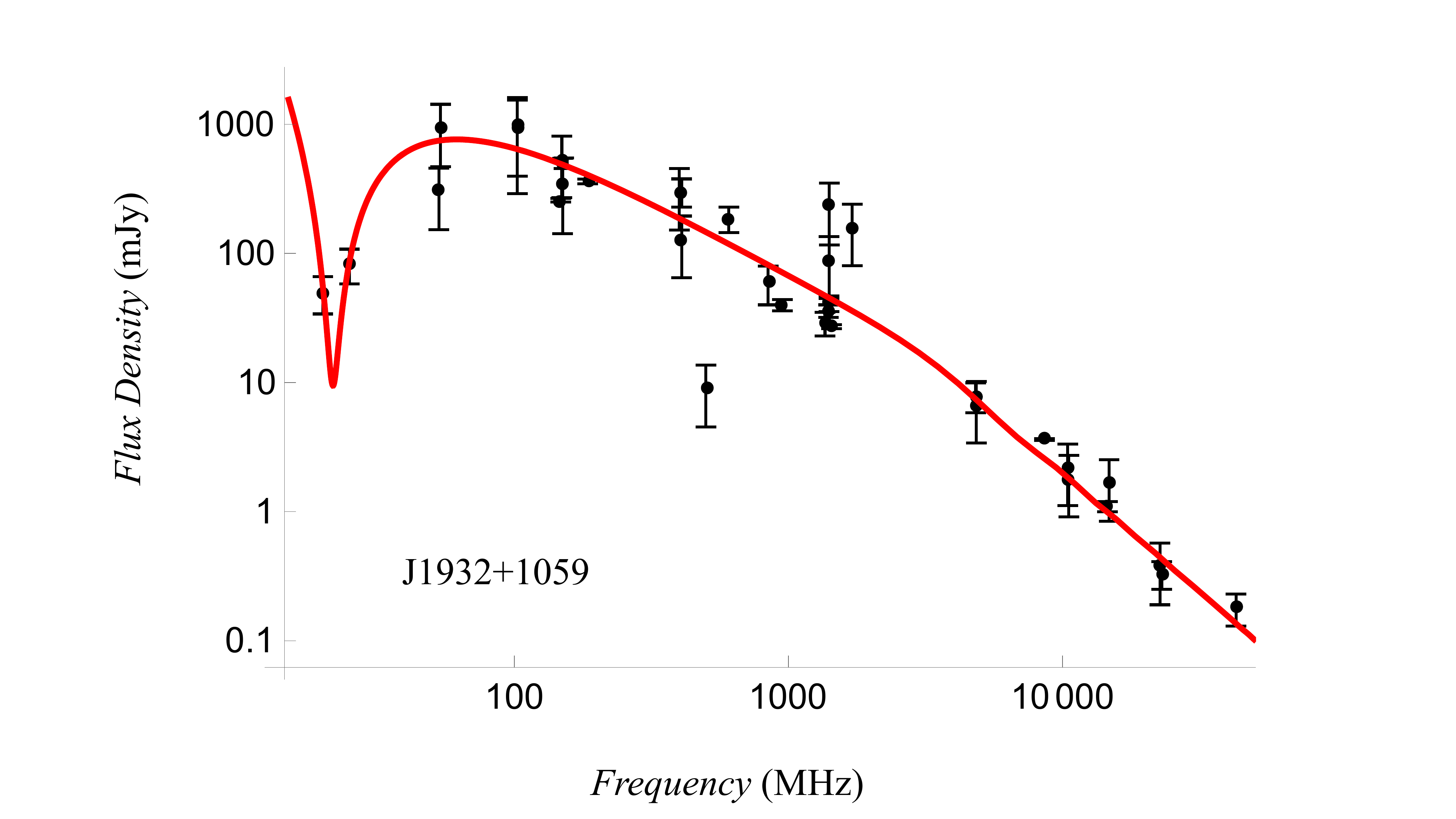}}
\caption{Radio spectrum of PSR J1932+1059 fitted with the spectral distribution function of the emission from its current sheet (see~\cite{Ardavan2024Radio} for other fitted radio spectra and for further details).}
\label{Radio3}
\end{figure} 

The enhanced focusing of the radiation with distance that is caused by this shortening of the separation between the turning points of a phase function results in a decay rate of the flux density with distance that is slower than that predicted by the inverse-square law.  Along colatitudes close to $\theta_{PS}$, the flux density $S$ of the radiation diminishes with increasing distance from its source as ${\hat R}_P^{-3/2}$ instead of ${\hat R}_P^{-2}$ (\cite{Ardavan2021}, Section 5.5).  This dependence of $S$ on ${\hat R}_P$, or equivalently $D$, is illustrated in Fig.~\ref{FH19} in the case where $\alpha=60^\circ$, $\theta_P=90^\circ$, $\kappa_{\rm u}=10^7$, and $D$ ranges from $1$ to $10^7$ kpc, i.e.\ from a galactic to a cosmological distance.  

As indicated by Fig.~\ref{FH6}b, the dependence $S\propto{\hat R}_P^{-3/2}$ of the flux density on distance holds within the interval $\vert\theta_P-\lim_{R_P\to\infty}\theta_{PS}\vert\le{\hat R}_P^{-1}$ of a critical latitude.  Further away from a critical latitude, e.g.\ at a larger interval $\vert\theta_P-\lim_{R_P\to\infty}\theta_{PS}\vert={\hat R}_P^{-\gamma}$ with $0\le\gamma\le1$, this dependence is given by
\begin{eqnarray}
&S&\big\vert_{\vert\theta_P-\lim_{R_P\to\infty}\theta_{PS}\vert={\hat R}_P^{-\gamma}}\nonumber\\*
&=&S\big\vert_{\theta_P=\lim_{R_P\to\infty}\theta_{PS}}{\hat R}_P^{-{\textstyle\frac{1}{2}}(1-\gamma)},\quad0\le\gamma\le1.\nonumber\\*
\label{E8}
\end{eqnarray}
Thus, the gradual change in the rate of decay of the flux density with distance from ${\hat R}_P^{-3/2}$ to ${\hat R}_P^{-2}$ away from the critical latitude $\lim_{R_P\to\infty}\theta_{PS}$ takes place over a latitudinal interval of the order of a radian.  Along a latitude $\theta_P$ that differs from $\lim_{R_P\to\infty}\theta_{PS}$ by ${\hat R}_P^{-1/2}$, for instance, $S$ decreases as ${\hat R}_P^{-7/4}$ with increasing ${\hat R}_P$. 

The violation of the inverse-square law encountered here is not incompatible with the requirements of the conservation of energy because the radiation process discussed in this paper is intrinsically transient.  Temporal rate of change of the energy density of the radiation generated by this process has a time-averaged value that is negative (instead of being zero as in a conventional radiation) at points where the envelopes of the wave fronts emanating from the constituent volume elements of the source distribution are cusped.  The difference in the fluxes of power across any two spheres centred on the star is thus balanced by the change with time of the energy contained inside the shell bounded by those spheres (see~\cite{Ardavan_JPP}, Appendix C, where this is demonstrated for each high-frequency Fourier component of a superluminally rotating source distribution).

Only the pulses, such as those depicted in Figs.~\ref{FH16}a and~\ref{FH18}b, that are created by the coalescence of two nearby stationary points disobey the inverse-square law.  The flux densities associated with the components of a pulse profile that arise from isolated stationary points of a phase, such as those depicted in Figs.~\ref{FH16}b and~\ref{FH18}a, decay like the flux density of a conventional radiation: as $R_P^{-2}$ with the distance $R_P$ from their source.  As we have already seen, the difference between the widths of these two types of pulse component at their peaks translate into a difference in the frequency bands at which they can be detected.  Hence, in cases where both types of pulse components are present in a given pulse profile, or along latitudinal directions where only the radio pulses are detectable, the high-frequency (gamma-ray, X-ray) and low-frequency (radio) luminosities of the observed pulses have to be estimated by using two different power laws for the decay of their fluxes with distance.

\begin{figure*}
\centerline{\includegraphics[width=15cm]{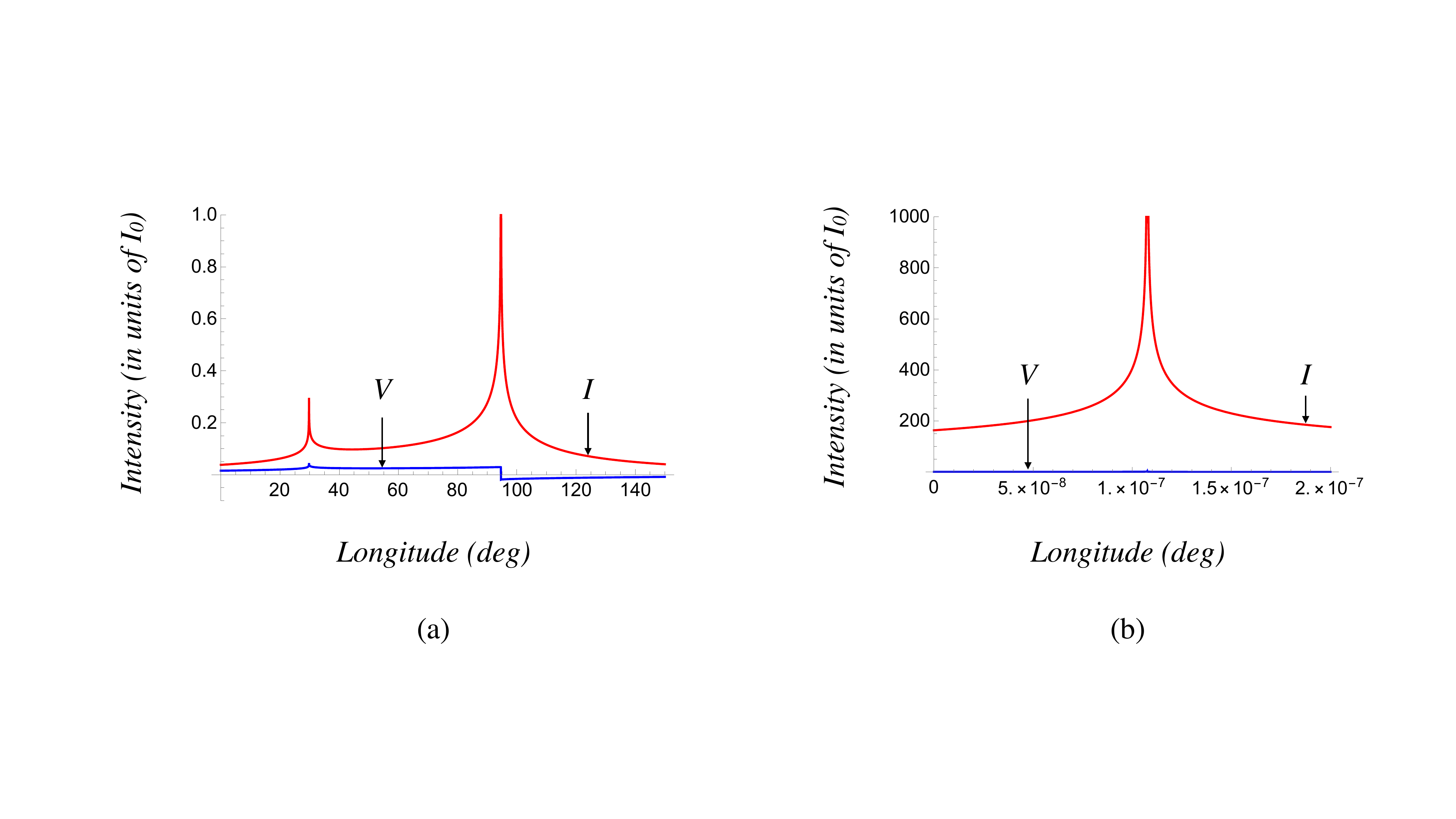}}
\caption{The Stokes parameters $I$ and $V$ for a three-component pulse profiltime-domain.  In this figure ${\hat R}_P=10^{13}$, $\theta_P=\lim_{R_P\to\infty}\theta_{PS}$, $\alpha=50^\circ$, $\kappa_{\rm u}=10^{10}$ and the origin of the horizontal axis in part (b) is shifted by $111.9695512^\circ$ relative to that of the horizontal axis in part (a).  It follows from the resolved widths of the peaks of these pulse components that while the two components shown in part (a) can be detected only in the radio band, the single component shown in part (b) is observable as a gamma-ray pulse. The radio and gamma-ray peaks of this pulse profile are thus separated by about $17^\circ$.}
\label{FH18}
\end{figure*} 

\begin{figure}
\centerline{\includegraphics[width=8cm]{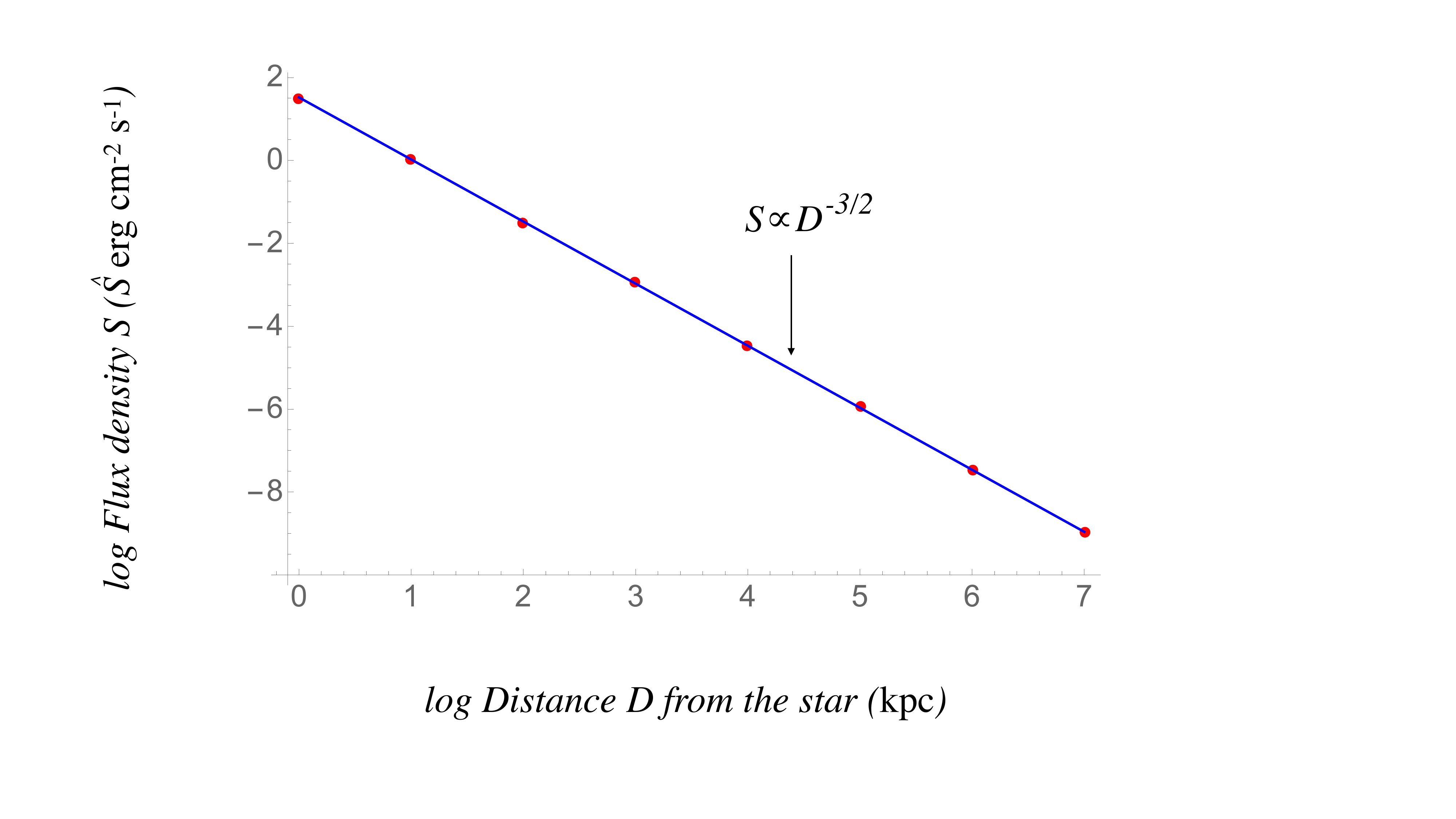}}
\caption{$\log(S/{\hat S})$ versus $log(D)$ for $\alpha=60^\circ$, $\theta_P=90^\circ$ and $\kappa_{\rm u}=10^7$.  The blue line with the slope $-3/2$ is the best fit to the red dots whose ordinates are determined by evaluating the expression in Eq. (\ref{E6}) for the flux density of the present radiation $S$ in units of ${\hat S}$.}
\label{FH19}
\end{figure} 

Luminosities of gamma-ray pulses are over-estimated when the decay rate of their flux density $S$ is assumed to obey the inverse-square law instead of $S\propto{\hat R}_P^{-3/2}$ by the factor ${\hat R}_P^{1/2}$.  The factor by which the luminosity of a $100$~ms gamma-ray pulsar at a distance of $2.5$~kpc is thus overestimated is approximately $4\times10^6$.  Once this is multiplied by the ratio $\sim1/30$ of latitudinal beam-widths of gamma-ray and radio pulsars (implied by the fraction of known pulsars that are detected in gamma-rays), we obtain a value of the order of $10^5$ for the overestimation factor in question: a result that implies that, contrary to the prevalent view~\cite{Lyne2012}, the range of values of the correctly-estimated luminosities of gamma-ray pulsars is no different from that of the luminosities of radio pulsars.  This result is confirmed by a nonparametric analysis of the data on the gamma-ray pulsars both in the Second Fermi-LAT Catalogue (see~\cite{Ardavan_Fermi}) and in the Fermi-LAT 12-Year Catalogue (see~\cite{Ardavan2023}).

The X-ray luminosities of magnetars are likewise over-estimated by the factor ${\hat R}_P^{1/2}$ when the decay rate of their flux densities is assumed to obey the inverse-square law.  The factor by which the luminosity of a $5$~s magnetar at a distance of $8$~kpc is over-estimated is approximately $10^6$.  Once this is multiplied by the ratio $\sim10^{-2}$ of latitudinal beam-widths of magnetars and radio pulsars (implied by the fraction of known neutron stars that are identified as magnetars), we obtain a value of the order of $10^4$ for the over-estimation factor: a result that implies that, contrary to the prevailing view~\cite{Kaspi2017}, the values of the correctly-estimated luminosities of magnetars are invariably lower than those of the spin-down luminosities of these objects, i.e.\ that the energetic requirements of magnetars are not different from those of spin-powered pulsars.  A nonparametric analysis of the data on the magnetars in the McGill Magnetar Catalog corroborates this result (see~\cite{Ardavan_magnetar}). 

The non-spherically decaying pulses that are created by the coalescence of two nearby stationary points of a phase function need not always be as narrow as those depicted in Figs.~\ref{FH16}a and~\ref{FH18}b.  The spectra of such pulses would peak in the radio band if the free parameter $\kappa_{\rm u}$ has lower values than those used in plotting Figs.~\ref{FH16}a and~\ref{FH18}b.  Given their limited latitudinal extent, the non-spherically decaying pulses generated by the current sheet of a neutron star are more likely to be detected (both in the low and high-frequency bands) when, as a result of movement (e.g.\ precession or a glitch) of the star's rotation or magnetic axes, the radiation beams embodying such pulses sweep past the Earth.  Using the decay rate $R_P^{-2}$, instead of $R_P^{-3/2}$, we would over-estimate the power emitted by the sources  of the bursts of radiation we would receive in this way by as large a factor as $10^{15}$ if the neutron stars that generate the bursts lie at cosmological distances.  

It is widely maintained that the powers emitted by distant sources of gamma-ray and fast radio bursts are by many orders of magnitude greater than that emitted by a pulsar~\cite{Kumar2015,Petroff2019,Zhang2023}.  The unquestioned assumption on which this consensus is based is that the radiation fields of all sources necessarily decay as predicted by the inverse-square law.  This assumption is brought into question by the results derived in~\cite{Ardavan2021} and described in this paper, however.  The putative difference between the energetic requirements of what are regarded to be different types of sources could arise from the difference in the latitudinal direction along which an obliquely rotating neutron star is observed. 

\section{Concluding remarks}
\label{sec:conclusion}

The following final remarks are in order: 
\begin{description}
\item[(i)] It is often presumed that the plasma equations used in the numerical simulations of the magnetospheric structure of an oblique rotator should, at the same time, predict any radiation that the resulting structure would be capable of emitting (see, e.g.~\cite{SpitkovskyA:Oblique,Contopoulos:2012}).  This presumption stems from ignoring the role of boundary conditions in the solution of Maxwell’s equations.  As we have already pointed out, the boundary conditions with which the structure of the pulsar magnetosphere is computed are radically different from the boundary conditions with which the retarded solution of these equations (i.e. the solution describing the radiation from the charges and currents in the magnetosphere) is derived (see Section~3 and the last paragraph in Section 6 of~\cite{Ardavan2021}).
 
 \item[(ii)] Thickness of the current sheet, which sets an upper limit on the frequency of the present radiation, is dictated by microphysical processes that are not well understood: the standard Harris solution of the Vlasov-Maxwell equations that is commonly used in analysing a current sheet is not applicable in the present case because the current sheet in the magnetosphere of a non-aligned neutron star moves faster than light and so has no rest frame.  Even in stationary or subluminally moving cases, there is no consensus on whether equilibrium current sheets in realistic geometries have finite or zero thickness~\cite{Klimchuk}.  The fact that the spectral distribution function of the emission from the current sheet of a non-aligned neutron star yields such good fits to both the radio and gamma-ray spectra of the pulsars shown in Figs.~\ref{Crab}, \ref{Radio1}, \ref{Radio2} and \ref{Radio3} lends support to treating the magnetospheric current sheet as volume-distributed but ultra-thin (see Section 4.7 of~\cite{Ardavan2021}).
 
\item[(iii)] In contrast to every one of the models currently considered in the literature on the subject~\cite{Beskin2018, Melrose2021} which are at least in part phenomenological, the all-encompassing body of explanations given here for the salient features of the observed emission from pulsars~\cite{Lyne2012, Abdo2013, Abdollahi2022} and for the putative energetic requirements of magnetars~\cite{Kaspi2017} and sources of fast radio bursts and gamma-ray bursts~\cite{Kumar2015,Petroff2019,Zhang2023} is a consequence purely of the basic equations that govern the magnetospheric structure of a neutron star and the emission of electromagnetic waves.
\end{description}

\bibliographystyle{plain}

\end{document}